\documentclass{emulateapj}
\usepackage[below]{placeins}
\usepackage{natbib}
\slugcomment{to be published by the Astrophysical Journal}

\def\gtorder{\mathrel{\raise.3ex\hbox{$>$}\mkern-14mu
    \lower0.6ex\hbox{$\sim$}}}
\def\ltorder{\mathrel{\raise.3ex\hbox{$<$}\mkern-14mu
    \lower0.6ex\hbox{$\sim$}}}
\def\hmpc{h^{-1} \mathrm{Mpc}}

\def\kmskpc{\ {\rm km~s^{-1} kpc^{-1}}}
\def\msun{M_\odot} 
\def\hmsun{$h^{-1} {\rm M_\odot}$}

\newcommand {\rs} {$R_{\rm s}$}
\newcommand {\rts} {$\tilde{R}_{\rm s}$}
\newcommand {\ms} {$M_{\rm s}$}
\newcommand {\mts} {$\tilde{M}_{\rm s}$}

\newcommand {\rvir} {$R_{\rm vir}$}
\newcommand {\mvir} {$M_{\rm vir}$}

\newcommand {\rvmax} {$R_{\rm vmax}$}
\newcommand {\mvmax} {$M_{\rm vmax}$}

\shorttitle{Dissecting Galaxy Formation}
\shortauthors{Romano-Diaz, Shlosman, Heller and Hoffman}

\begin{document}

\title{Dissecting Galaxy Formation:\\
I.\/ Comparison Between Pure Dark Matter and Baryonic Models}

\author{ 
Emilio Romano-D\'{\i}az\altaffilmark{1},
Isaac Shlosman\altaffilmark{2,3},
Clayton Heller\altaffilmark{4},
Yehuda Hoffman\altaffilmark{5}
}
\altaffiltext{1}{
Department of Physics and Astronomy, 
University of Kentucky, 
Lexington, KY 40506-0055, 
USA
}
\altaffiltext{2}{
JILA,
University of Colorado,
Boulder, CO 80309-0440,
USA
}
\altaffiltext{3}{
National Institute of Standards and Technology,
Boulder, CO 80305-3328,
USA
}
\altaffiltext{4}{
Department of Physics, 
Georgia Southern University, 
Statesboro, GA 30460, 
USA
}
\altaffiltext{5}{
Racah Institute of Physics, Hebrew University; Jerusalem 91904, Israel
}

\begin{abstract}
We compare assembly of dark matter (DM) halos with and without baryons from 
identical initial conditions, within the context of 
cosmological evolution in the $\Lambda$CDM WMAP3 Universe (baryons$+$DM, 
hereafter BDM model, and pure DM, PDM model). In 
representative PDM and BDM models we find that baryons contribute decisively
to the evolution of the central region, leading to an isothermal DM cusp, and 
thereafter to a flat DM density core --- the result of heating by dynamical 
friction of the DM$+$baryon substructure during a quiescent evolution epoch. 
This process ablates the cold gas from an embedded disk, cutting the star 
formation rate by a factor of 10, and heats up the spheroidal
gas and stellar components, triggering their expansion. The substructure
is more resilient to the tidal disruption in the presence of baryons. The disk 
which formed from inside-out as gas-dominated, is transformed into an 
intermediate Hubble type by $z\sim 2$ and to an early type by $z\sim 0.5$, 
based on its gas contents and spheroidal-to-disk stellar mass ratio. We find 
that only a relatively small $\sim 20\%$ fraction
of DM particles in PDM and BDM models are bound within the radius of maximal 
circular velocity in the halo, slightly less so within halo characteristic 
radii --- most of the DM particles perform larger radial excursions. The DM 
particles are unbound to the cusp region. 
We also find that the fraction of baryons within 
the halo virial radius somewhat increases during the major mergers and 
decreases during the minor mergers. The net effect appears to be negligible
--- an apparent result of
our choice of feedback from stellar evolution. Furthermore, we find
that the DM halos are only partially relaxed beyond their virialization.
While the substructure is being tidally-disrupted, mixing of its debris
in the halo is not efficient and becomes even less so with $z$. The 
phase-space correlations (streamers) formed after $z\sim 1$ will survive 
largely to the present time --- an important implication for embedded disk 
evolution.  
\end{abstract}

    
%-------------------------------------------------------------------
\section{Introduction}
\label{sec:intro}
 
Luminous parts of galaxies are known to form inside dark matter (DM)
halos (e.g., White \& Rees 1978).  Within the CDM paradigm, baryons
are well mixed with DM initially.  The collapse and virialization of
DM is accompanied by infall of the baryonic matter. Dissipation
results in the partial separation between these components --- a
process which can be slowed down by gas-to-stars conversion in
ellipticals and by the angular momentum barrier in disk galaxies. This
separation can lead to a number of processes which amplify the angular
momentum and energy transfer between DM and baryons when the latter
accumulate in the central region of the potential well. Understanding
these processes is of prime importance because galaxy evolution
is largely driven by mass, energy and angular momentum flows.

Within the framework of the CDM WMAP3 cosmology 
(Spergel et al. 2003, 2007), the baryons account for non-negligible,
$\sim 17\%$, mass fraction. It is important to ask, therefore, what
effect the baryons have on the DM halo evolution and vice versa. The
main issue which we attempt to address here and in the companion paper
(Romano-Diaz et al., in preparation, Paper~II) is to what degree the
DM halo, its buildup and relaxation, as well as
properties and evolution of the surrounding substructure are affected by the
baryons. Consequently, we explore in depth the baryon aspects of DM
dynamics in the halo. We follow the baryon evolution as gas and stars
which subject the system to energy and momentum feedbacks. Contrary to
the statistical approach prevailing in the field, which aims at typical
properties of growing halos, we focus on a small number of models
which start from identical initial conditions. 
The models differ only by the presence or absence of baryons, hereafter
baryons$+$DM (BDM model) and pure DM (PDM). As a prototype, we choose
a massive Milky Way-type halos which have experienced an early epoch of
major mergers, but remained isolated since $z\sim 1.5$.

The scope of baryon--DM interactions in context of galaxy evolution
has been explored directly and indirectly over some time (e.g.,
Shlosman 2007 and refs.  therein). For example, the halos appear to
have small angular momentum compared to baryons in disks.  As a
result, the angular momentum can naturally flow from the disk to the
surrounding halo, especially if triggered by lack of axial symmetry
(e.g., Lynden-Bell \& Kalnajs 1972; Tremaine \& Weinberg 1984;
Debattista \& Sellwood 1998; Athanassoula \& Misiriotis 2002; 
Athanassoula 2002; Martinez-Valpuesta, Shlosman \& Heller 2006).  
Hence, triaxial shapes of DM halos consistently
obtained in pure DM cosmological simulations (e.g., Allgood et
al. 2006) are expected to facilitate such an angular momentum transfer
from baryons, despite the latter ability to lower this triaxiality
(Kazantzidis et al. 2004; Berentzen \& Shlosman 2006).

One cannot understand the full extent of baryon-DM interactions
without in-depth analysis of the dynamical state of DM in a pure
DM halo. This, however, has encountered a number of difficulties.  We
do not understand the origin of some of the seemingly most robust
parameters defining the halo properties, such as the density profiles
(e.g., Navarro, Frenk \& White 1997, hereafter NFW; El-Zant 2008), the
phase-space density profiles (e.g., Taylor \& Navarro 2001; Hoffman et
al. 2007), and the density slope -- velocity anisotropy relation
(e.g., Hansen \& Moore 2006; Barnes et al. 2007; Zait, Hoffman \&
Shlosman 2008), and others. Which parameters are the most important
ones in characterizing the growing halo is currently debatable. What
is the relevance of the characteristic (NFW) radius \rs? As this 
parameter is
obtained by spherically-symmetrizing the halos, what significance
it has for arbitrary-shaped halos is not clear.
 
Furthermore, while an assembling halo goes through the virialization
process, its relaxation both in the configuration and velocity spaces
is not well understood.  The existence of substructure (e.g.,
subhalos) within the context of the hierarchical clustering is
inherent and subject to investigation (e.g., Tormen et al. 1998;
Ghigna et al. 1998; Moore et al. 1999; Klypin et al. 1999a,b; Gao et
al. 2004a,b; Gnedin et al. 2004; Reed et al. 2005; Diemand et al. 
2008; Springel et al. 2008). This includes the
tidal disruption of subhalos and their subsequent mixing with the
background material. It is expected that even after full tidal
dissolution of subhalos, the associated currents (streamers) will 
persist for a longer time period. Both subhalos and streamers
when penetrating the central region can affect the disk evolution
in various ways which can be, in principle, observable. Hence,
it is important to understand the role of baryons in the subhalo
evolution --- if baryons can substantially modify the subhalos
and their remnants, ultimately this may affect the disk evolution
as well.

The effect of baryons on the evolution of the DM substructure was
considered recently by Weinberg et al. (2008). They find that the DM
dictates the galaxy clustering while the baryons affect the
small-scale DM distribution.  The fate of the subhalos appears to be
most heavily influenced in the denser regions, where the baryons
enhance the binding of subhalos.  While we defer the comparison
between the substructure evolution with and without baryons to
Paper~II, various aspects of prime halo --- substructure interactions
are analyzed here as well. We confirm some of the evolutionary trends
addressed by Weinberg et al. (2008).  Furthermore, due to the superior
resolution of our numerical simulations on a galactic scale, e.g., 
nearly three orders of
magnitude in mass resolution, we are able to zoom into specific
dynamical processes which accompany the prime halo buildup, the
substructure evolution, and halo relaxation processes in the presence
of baryons.

While we adopt the current characterization of DM halos in terms of
the spherical overdensity, i.e., the NFW profile and the 
characteristic radius \rs{} for the pure DM simulations, we also attempt 
to break away from this description in favor of arbitrarily shaped 
halos. Furthermore, the NFW density profile does not fit the 
DM distribution in the presence of baryons (e.g., Romano-Diaz et al.
2008c), especially in the inner halo and so we drop this approach
altogether. 

This paper is
structured as follows. \S2 deals with numerics, initial conditions and
related issues. \S3 and \S4 present results of numerical simulations and
their analysis, and \S5 describes
the global baryon evolution, while the issues of disk evolution will
be discussed elsewhere. Discussion and conclusions follow in \S6.
 
%%%%%%%%%%%%%%%%%%%%%%%%%%%%%%%%%%%%%%%%%%%%%%%%%%%%%%%%%%%%%%%%%%%
\section{Numerics and Initial Conditions}

Numerical simulations have been performed using the parallel version of
FTM-4.5 hybrid $N$-body/SPH code (e.g., Heller \& Shlosman 1994;
Heller, Shlosman \& Athanassoula 2007). The total number of DM
particles is $N\approx 2.2\times 10^6$ and of the SPH particles is
$4\times 10^5$. The gravitational forces are computed using the falcON
routine (Dehnen 2002) which is about ten times faster than
optimally-coded Barnes \& Hut (1986) tree code and and scales as
\cal{O}(N). The tolerance parameter $\theta$ was fixed at 0.55.  The
gravitational softening is $\epsilon=500$~pc for the DM, stars and
gas. The mass density and gravitational potential of the softening 
kernel are given as (Walter Dehnen, private communication)

\begin{equation}
\rho(r) = \frac{15}{8\pi} \frac{\epsilon^4}{(r^2+\epsilon^2)^{7/2}}
\label{eq:rho_soft}
\end{equation}

\begin{equation}
\Phi(r) = -\frac{G}{\sqrt{r^2+\epsilon^2}}
        \left[1 + \frac{\epsilon^2}{2(r^2+\epsilon^2)}\right].
\end{equation}
The density in eq.~\ref{eq:rho_soft} falls off faster than a Plummer 
sphere at large 
radii, i.e., $r^{-7}$ vs. $r^{-5}$, which avoids the force bias inherent 
to Plummer softening (Dehnen 2001). 

We use the vacuum boundary conditions and perform simulations
with physical coordinates.  The cosmological constant is introduced by
an explicit term in the acceleration equation. We assume the
$\Lambda$CDM scenario with WMAP3 parameters: $\Omega_{\rm m}=0.24$ and
$\Omega_\Lambda=0.76$ and $h=0.73$, where $h$ is the Hubble constant
in units of $100~{\rm km~s^{-1}~Mpc^{-1}}$. The variance
$\sigma_8=0.76$ of the density field convolved with the top-hat window
of radius $8h^{-1}$~Mpc was used to normalize the power spectrum. The
conservation of the total angular momentum and energy within the
computational sphere in the PDM simulations has been followed and is
within $\sim 0.01\%$ and $\sim 1\%$ respectively. The evolution of
various parameters characterizing the DM and baryons has been
followed in 1,000 snapshots, linearly spaced in the cosmological
expansion parameter $a$. In these simulations we have switched off
the external UV background.
 
\subsection{Star Formation and Feedback Processes}

We model star formation (SF) processes and associated feedback as
described in Heller \& Shlosman (1994) and Heller et al. (2007),
which should be consulted for details. For SF to occur 
the local gas must meet several conditions, including being 
Jeans-unstable. Feedback from OB stellar winds and supernovae (SN)
Type~II is accounted for by injecting energy from SN
and stellar winds into the $N_{\rm SF} = 16$ surrounding gas 
particles. The time step of such ``active''
stellar particles (and of all the gas particles) is restricted in 
order to properly resolve the feedback timescale. 
The radiative cooling of $N_{\rm SF}$ gas particles in the vicinity 
of the ``active''  stellar particles is temporarily disabled when 
receiving the energy from a stellar particle.
A fraction of this energy is thermalized and deposited in the gas in 
the form of a thermal energy, then converted to kinetic energy 
through the equations of motion. This method is preferable over 
injecting a fraction of the stellar energy directly in the form of 
a kinetic energy. 

The resulting SF rate is that of the 
Schmidt-Kennicutt law (e.g., Kennicutt 1998)
and depends on the adopted fudge factors which determine the stellar
energy feedback, $\epsilon_{\rm SF}$ (i.e., the fraction of the
thermalized energy), the threshold for the SF as a fraction of the
background gas density, $\alpha_{\rm crit}$, and the collapse
time of a cloud in terms of the local dynamical time, $\alpha_{\rm
  ff}$. These factors are fixed at $\epsilon_{\rm SF}=0.3$,
$\alpha_{\rm crit}=0.5$ and $\alpha_{\rm ff}=1$ following Heller et
al. (2007).
 
We introduce the probability that a gas particle of mass $m_{\rm g}$
produces a stellar particle of mass $m_{\rm s}$ during a given timestep and
multiple generations of stars are allowed to form from each gas particle. 
A fraction 0.4 of the stellar mass is instantaneously recycled to the
parent gas particle. The evolution of gas metallicity is followed and
the fraction of massive stars that lead to the OB stellar winds and SN
is calculated from the Salpeter IMF.

\begin{figure*}
\begin{center}
\includegraphics[angle=0,scale=0.9]{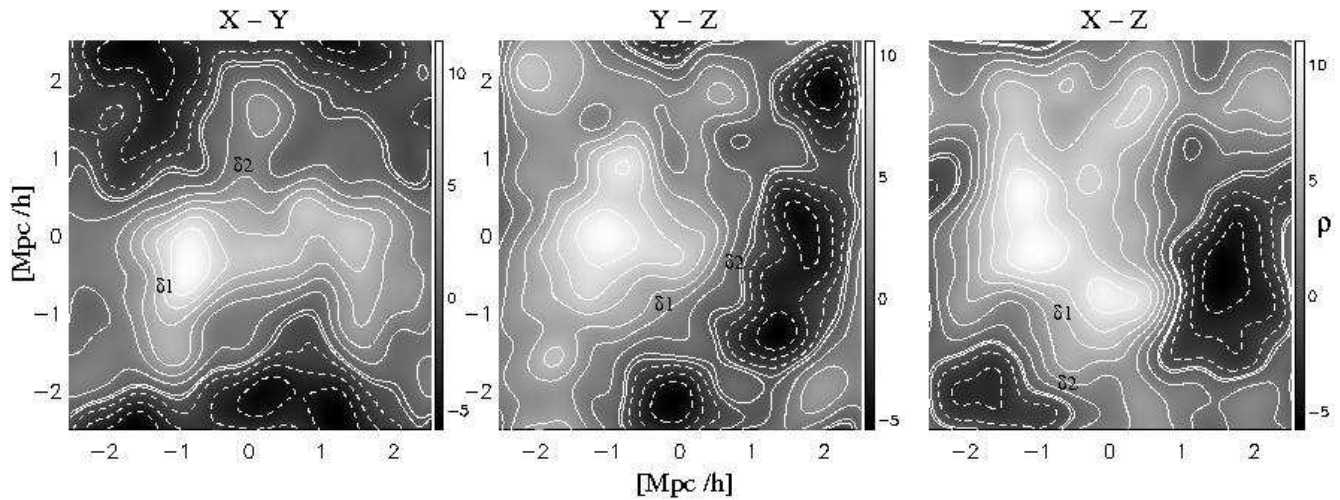}
\end{center}
\caption{Initial conditions: The linear over-density field presented on 
the three principal planes of the simulation, going through the center 
of a box of co-moving $5h^{-1}$~Mpc on the side. This field
is normalized by its present epoch value and is presented with a 
Gaussian smoothing on a mass scale of $10^{10}$\hmsun. The thick solid 
line corresponds to the over-density $\delta=0$, and solid (dashed) lines 
correspond to positive (negative) perturbations. The imposed constraints 
are $\delta_1=3.0$ on a mass scale of $M= 10^{12}$\hmsun{} and $\delta_2=0$ 
on a mass scale of $M=5\times 10^{13}$\hmsun. Both are imposed at the center 
of the box.
}
\end{figure*}

The thermal balance in the gas is calculated using the energy
equation. Adiabatic, viscous and radiative processes are included in
the gas heating and cooling. The fractions of H$^+$, He, He$+$,
He$^{++}$ and e$^-$ are calculated in tandem with the mean molecular
weight as a function of density and temperature, assuming
optically-thin primordial composition gas.

\subsection{Initial Conditions: Constrained Realizations Method}
 
We use the method of Constrained Realizations 
(CRs, Bertschinger 1987; Hoffman \& Ribak 1991; van de Weygaert \&
Bertschinger 1996) and follow the prescription of Hoffman \& Ribak (1991)
to build the initial conditions within a restricted cubic volume of space
with sides
$L=8 \hmpc$ in the $\Lambda$CDM cosmology, where a sphere of
$5h^{-1}$~Mpc is carved out and evolved from $z=120$. The constructed
Gaussian fields obey a set of constraints of arbitrary amplitudes
and positions.  The CR algorithm is exact, involves no iterations and is
based on the property that the residual of the field from its mean is
statistically independent of the actual numerical value of the
constraints (see also Romano-Diaz et al. 2006, 2007). This method
allows tailoring of the initial conditions to explore how fundamental
characteristics of the structure formation history affect the resulting
properties of DM and baryons within the computational cube.

A series of linear constraints have been applied on the initial
density field. Each of the constraints represents the value of the
initial density at various locations and evaluated with different
Gaussian smoothing kernels --- their width fixed in order to encompass
a mass $M$ (the mass scale on which the constraints are imposed).  The
designed models are based on two constraints. The first one has the mass
of $1.0\times 10^{12}~h^{-1}M_\odot$. This implies a $\delta = 3$
overdensity, which constitutes a $2.5\sigma$ perturbation and was
imposed on a $256^3$ grid. It is projected to collapse at 
$z_{\rm c}\sim 1.3$, based on the top-hat model, and is embedded in a region
(2nd constraint) corresponding to the mass of $5\times
10^{13}~h^{-1}M_\odot$ in which the over-density is zero,
corresponding to the unperturbed universe. The random component in CRs
favors formation of similar structures which leads to major mergers.

The total mass inside the computational sphere is $\sim 6.1\times
10^{12}~h^{-1}\msun$. To introduce the baryons,
we have replaced some DM particles on the initial conditions grid
by a baryon (i.e., Smooth Particle Hydrodynamics, hereafter SPH) particles, 
so that $\Omega_{\rm m}$ stays the same for pure DM and baryonic 
perturbations. The initial masses of DM and SPH particles are the same,
$2.78\times 10^6~{\rm M_\odot}$.

\begin{figure*}
\begin{center}
\includegraphics[angle=0,scale=0.9]{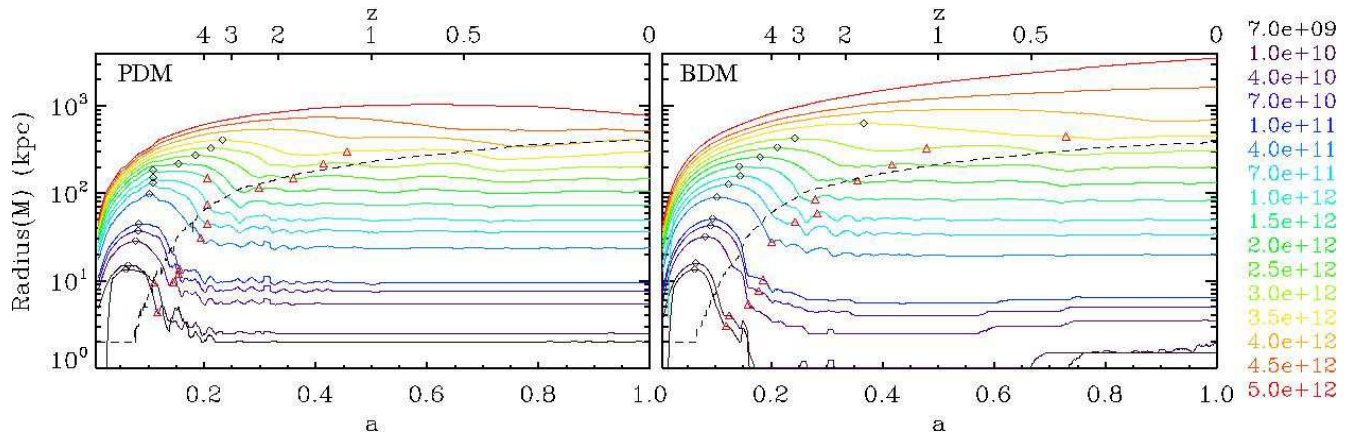}
\end{center}
\caption{DM halo buildup in the PDM (left) and BDM (right)
simulations: evolution of radii containing the fixed masses. 
The dashed lines display \rvir, open diamonds --- the turnover  
time for each mass, and triangles --- the collapse time measured as 
twice the turnover time. \mvir{} at $z=0$ are $4\times 10^{12} 
\msun$, and the \rvir{} radii in the PDM and BDM are nearly identical, 
$\sim 400$~kpc.  }
\end{figure*}

\begin{figure}
\begin{center}
\includegraphics[angle=0,scale=0.33]{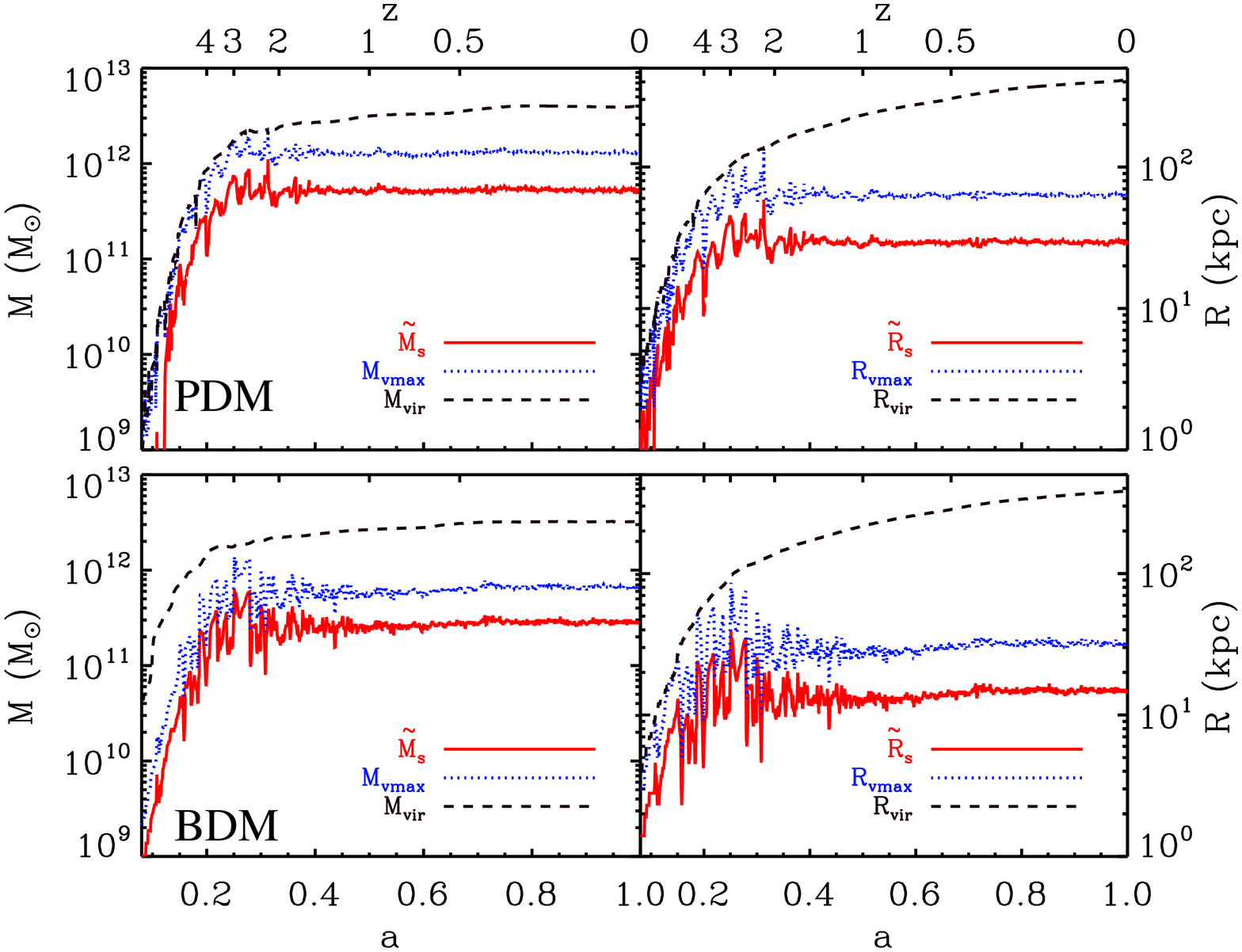}
\end{center}
\caption{Evolution of the characteristic masses (in $\msun$) and radii (kpc) 
for the prime halo in PDM (upper) and BDM (lower) models. \mts{} and \rts{}
(red), \mvmax{} and \rvmax{} (blue), and \mvir{} and \rvir{} (black). 
 }
\end{figure}

%%%%%%%%%%%%%%%%%%%%%%%%%%%%%%%%%%%%%%%%%%%%%%%%%%%%%%%%%%%%%%%%%%%%
\section{Results: PDM and BDM halos buildup}

\subsection{Primary Halo Characteristic Radii, Radial 
Density Profiles and Concentrations}

We use the DM mass
density peak in order to find the prime halo (by means of the HOP algorithm,
Eisenstein \& Hut [1998]). The corresponding subhalo finding 
algorithm is described in Paper~II. The subhalo tidal radii are described 
in \S3.6. The density peaks of all three DM and baryonic components 
may not coincide among themselves. As we follow
the buildup of the prime halo, we sample it at
characteristic radii defined below.   
The outermost radius describes the region that has just virialized.
Smaller characteristic radii depend more on the density peak
properties, where virialization is more complete. 
 
The initial (linear) density field corresponds to a filament running
across the computational box and having a `banana' shape (Fig.~1).
The prime halo assembly is shown in Fig.~2. The major mergers epoch
ends at $t\sim 4.5$~Gyr, which corresponds to $z\sim
1.5$, in agreement with the prescribed $z_{\rm c}$.

The halo virial radius is defined here in the context of the spherical
top-hat collapse model, \mvir = 4/3$\pi \Delta(z)\rho(z)$\rvir$^3$,
where $\Delta(z)$ (Bryan \& Norman 1998) is the critical overdensity 
at virialization.
It signifies the part of the flow which is virialized, at least
partially.  As seen in Fig.~2, the outermost spherical
surface contains DM mass which by $z=0$ is about \mvir $\sim
3.5\times 10^{12}~\msun$ (i.e., $\sim 3.9\times 10^{12}~\msun$ 
and $\sim 3.2\times 10^{12}~\msun$, in PDM and BDM models, respectively)
and has \rvir$\sim 400$~kpc (i.e., 412~kpc and 385~kpc, respectively). 
Hence baryons
have a minimal impact on the halo overall size and mass in these
simulations.  During the major merger epoch, \rvir{} in Fig.~2
accurately follows the line of the open triangles which delineates the
double of the turnover time, i.e., the collapse time for this mass.
At later times, the actual collapse time is slightly longer than the
top-hat estimate. Most of the mass shows only mild
virial oscillations after they enter \rvir. We note that \mvir\
curve is an over-estimate during the major merger epoch.
\rvir\ of the prime halo and neighboring massive subhalos overlap
because of the initial conditions, and the mass of the subhalos
is added to the prime halo in this case.

The virial masses of the prime halos in PDM and BDM models grow in tandem
most of the time. However, around $z\sim 4$, the PDM halo becomes more
massive by a factor of 5 for a brief period of time. We do not attribute
any real significance to this event which is related to a slightly
shifted merging history.

The radial DM density profiles in the PDM model are well fitted by
the NFW profile and define the characteristic radius \rs\ where the
log$\rho-$log$R$ slope is --2. A comparison between the PDM and
BDM halo density profiles reveal substantial evolutionary differences, 
even in the 
early stages, at $z\sim 5$. At low $z$, the DM density profiles are 
similar outside $\sim 20$~kpc, the PDM profile being
slightly higher than the BDM one (Romano-Diaz et al. 2008b).  Inside
this radius, the BDM density profile is steeper and takes over. 
Romano-Diaz et al. (2008b) have analyzed the evolution of
this region and shown that after $z\sim 4$ the DM in the BDM acquires
the slope of an isothermal distribution, $-2$, in log$\rho-$log$R$
within the central
$\sim 15$~kpc --- the result of an adiabatic contraction. This {\it
isothermal} cusp is gradually leveled off after $z\sim 1$ by the
heating action of accreting subhalos. The flat core forms within the
inner $2-3$~kpc. We do not pursue this issue 
further here and only comment that the flattening of the isothermal 
cusp cannot be related to the
finite resolution of the code as it produces less than $5\%$
difference between the gravitationally softened force and the exact
Newtonian force of point masses at 1~kpc from the center.
Hence, the BDM density profile cannot be fitted
by the NFW profile around \rs{} and at smaller radii. We, therefore,
define an alternative radius which samples this region. 

The radius of a maximal circular velocity, \rvmax, in the DM halo has a 
universal character when the NFW fit is used, and 
\begin{equation}
\gamma\equiv\frac{R_{\rm s}}{R_{\rm vmax}} \sim 0.46. 
\end{equation}
This radius provides a robust
sampling of inner halo properties. We define a characteristic radius \rts\ 
which is a constant fraction $\gamma=0.46$ of \rvmax{}
and apply it instead of \rs{} to the BDM and PDM models. In the latter
case, \rts$\equiv$\rs, by definition. The three radii, 
\rvir, \rvmax{} and \rts{} are utilized by us to
sample the outer and inner halos. Quantities defined
at \rts{} will bear {\it tilde}.

The final \rts{} reach $\sim 28$~kpc and $\sim 15$~kpc, 
while \rvmax{} becomes 62~kpc and 32~kpc, in PDM and BDM,
respectively. The 
growth of \rts{} and \rvmax{} is essentially terminated after
the epoch of major mergers, but \rvir\ continues to grow albeit at a
slower pace (Fig.~3). This agrees well with earlier results (e.g.,
Wechsler et al. 2002, Romano-Diaz et al. 2006, 2007; Diemand, Kuhlen
\& Madau 2007).
We note that quantities defined within \rts{} and \rvmax{} typically show
a very similar temporal behavior, e.g., the mass within these radii.

\begin{figure}
\begin{center}
\includegraphics[angle=0,scale=0.4]{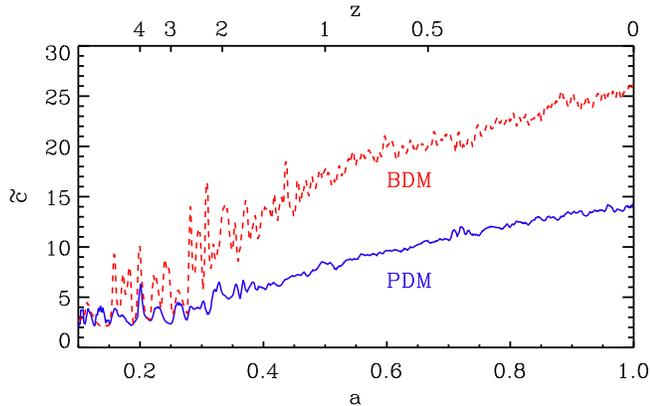}
\end{center}
\caption{Evolution of the concentration paramater
$\tilde{c}$ in PDM and BDM models.}
\end{figure} 
 
The concentration parameter, $\tilde{c} \equiv $\rvir/\rts, reveals the
differences in the evolution of the outer and inner halos. It shows a
substantial departure of PDM from the BDM model (Fig.~4). The source of
this difference is the adiabatic contraction of the DM which results
in a more concentrated halo. 

\begin{figure}
\begin{center}
\includegraphics[angle=0,scale=0.4]{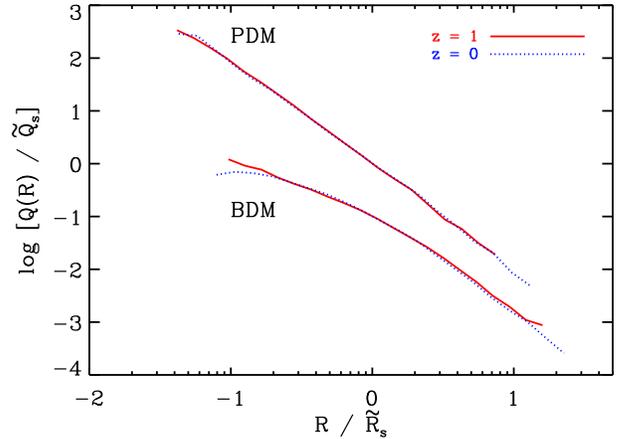}
\end{center}
\caption{The DM phase-space density, $Q(R)$ for PDM and BDM halos.
The x-axis was normalized by \rts{} at $z=1$ and 0, and 
y-axis by the value of $Q$ at \rts, 
$\tilde{Q}_{\rm s}$, at these $z$. The BDM curves are displaced by
0.1 vertically down for clarity.}
\label{fig:phase_den}
\end{figure}

Determining the halo's center-of-mass (CoM) is not trivial, especially
when baryons are present in the form of stars and gas, in addition to
the DM.  The main complication comes from the
assembling halo being an open system, both during the quiescent accretion
and major merger phases. 
The CoM normally does not coincide with the density peak. Moreover,
the density peaks of all three DM and baryonic components may not
coincide among themselves, especially because the halo is not fully
relaxed and is found only in a rough virial equilibrium.
During a merger
event, the position of the halo CoM jumps nearly discontinuously, when
\rvir{} is penetrated, but dynamics of the cold baryons (i.e., the
disk) which are assembled within the central 10~kpc -- 20~kpc is unaffected,
unless there is a direct intrusion into this central region before the
intruder is largely dissolved. In this latter case the disk dynamics
is affected dramatically. As we shall see, this is not limited to the
merger epoch only, and is important during the subhalos influx after
$z\sim 1$ as well.  

\subsection{Primary Halo in the Phase-Space}

We have compared the phase-space density, $Q(R)$ in the PDM and BDM
halos (Fig.~\ref{fig:phase_den}). While we confirm that for the PDM 
it can be fitted by the
power law  $Q(R) = \tilde{Q}_{\rm s} (R/R_{\rm s})^{-\beta}$, with
$\tilde{Q}_{\rm s}\equiv \tilde{\rho}_{\rm s}/\tilde{\sigma}_{\rm s}^3$, 
where $\tilde{\sigma}_{\rm s}$ is the 1-D radial DM dispersion 
velocity at \rts{} and $\beta\sim 1.95$ in the 
quiescent epoch (e.g., Hoffman et al. (2007). This dispersion velocity
has been calculated using the SPH kernel with 64 neighbors. However, 
the BDM halo
cannot be fit by this law after $z\sim 4$. The difference between
the models is most visible within the central $\sim 30$~kpc,
where the BDM $Q(R)$ starts to flatten. This process accelerates
after $z\sim 0.5$, when $Q$ develops a flat core within the central
few kpc. 

\subsection{Primary Halo Shapes}

We follow Heller et al. (2007) in calculating the DM halo
shapes. After substracting the halo COM velocity, we compute the
eigenvalues of the moment of inertia tensor for the DM mass within a
specified radius. Next, we determine the semi-axes $a>b>c$ of a
uniform spheroid using these eigenvalues.  The axes ratios are used to
characterize the halo shapes as defined by
$b/a=\sqrt{(e_1-e_2+e_3)/\Delta}$, and
$c/a=\sqrt{(e_1+e_2-e_3)/\Delta}$, where $\Delta=e_2-e_1+e_3$ for the
eigenvalues $e_3>e_2>e_1$. The axial ratios have been computed in
two ways, as a function of time and as a function of $R$. In the
former case, we calculate an integral value of $b/a$ and $c/a$
which represents the entire DM within \rvir. In the latter case,
we use the DM shells and look at the radial distribution of their axial
ratios. 

We confirm that the baryons do have a substantial effect on the halo
axial ratios, after a detailed comparison as shown by Dubinski (1994), 
Kazantzidis et al. (2004) and Berentzen \& Shlosman (2006). The ratios 
$b/a$
and $c/a$ are decreasing until the end of the major merger epoch at
$a\sim 0.4$, except for $b/a$ of the inner layers in the BDM halo. 
Thereafter, the halos
experience a gradual decrease in their equatorial ellipticity,
$\epsilon_\rho=1-b/a$, and flatness, $f_\rho=1-c/a$, with time, at all
radii. For the outer radii, this trend is more obvious and the 
{\it gradient} of triaxiality is slowly erased, but the outer layers
remain always more triaxial.  This is more or less in line with Heller
et al. (2007).  At $z=0$, the globally averaged axial ratios are
$b/a\sim 0.8-0.9$ and $c/a\sim 0.7-0.8$ for the PDM model. For the 
BDM, these
ratios are larger, $b/a\sim 0.9-0.95$ and $c/a\sim 0.8-0.9$ at the
end. The same trend is observed for the radial profiles of the axial
ratios. The halos appear slightly more elliptical at the intermediate
radii and nearly axisymmetric in the central 10~kpc. The $c/a$ seems
to be independent of $r$, except in the central 10~kpc where it tends
to unity.

To summarize, while the outer layers of PDM and BDM halos evolve in 
a similar
fashion, i.e., become more triaxial during the major merger epoch, and
less triaxial during the subsequent smooth accretion permeated by
minor mergers, the inner layer of the BDM halo starts to lose its
$\epsilon_\rho$ early, after the initial collapse. The outer halos
appear prolate at all times, while inner halos are nearly oblate at
lower $z$. We note, that while relative differences between the PDM
and BDM simulations appear robust, they must be tested also against
halos with different merger histories.

\subsection{Primary Halo Figure Tumbling and Angular Momentum}

Calculating the halo shape (\S3.3) and following the orientation of
the halo major axis, we find that the halo figures in PDM and BDM
simulations are essentially nonrotating (Fig.~6), based on the
temporal resolution of $\sim 900$ snapshots. Rather the halos librate
around a fixed direction by $\pm 10^\circ$. Our dense temporal
coverage allows the disentanglement of the effects of moving 
substructures (e.g.,
subhalos and tidal streams) which can otherwise introduce aliases and
result in erroneous estimates for the halo figure rotation. After the
epoch of the major mergers, and especially toward the end of the
simulations, the triaxiality gradients between the inner and outer
halo shells are substantially diminished,
hence the effect of the outer shells on the overall shape is not so 
pronounced.  

The difference between the exceedingly slow halo tumbling found in the
previous models and the non-tumbling halos in the present simulations
is {\it dynamically} unimportant. In the current models, the orientation 
of the halo major axis is nearly fixed in time with respect to the
orientation of the main DM filament --- the consequence of
gravitational torques from the filament which is able to focus the
smooth and clumpy accretion onto the halos. While the torque on the
filament itself from the large-scale structure is ignored here,
if this torque is passed onto the halo, the resulting halo tumbling will
be still dynamically insignificant.

\begin{figure}[ht!!!!!!!!!!!!!!!!!!]
\begin{center}
\includegraphics[angle=0,scale=0.4]{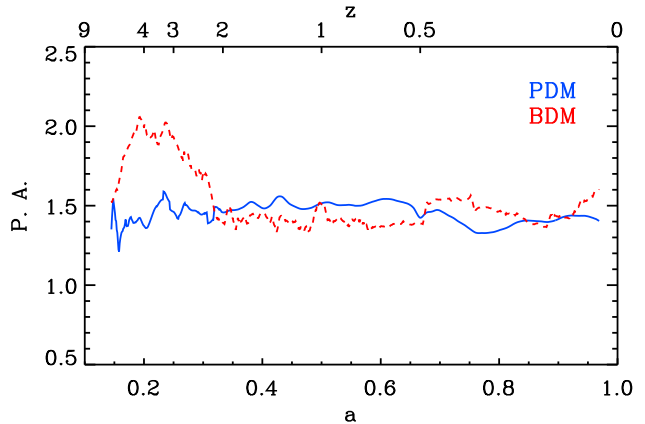}
\end{center}
\caption{The position angle (P.A., in radians) of the major 
axis of DM halo in PDM (solid, blue) and BDM (dashed, red) models. 
Based on $\sim 900$ snapshots.  
}
\end{figure}

While the prime halo axes librate around a preferred direction,
its internal circulation angular momentum shows much more
evolution. Here we refer to the momenta of DM, unless stated
otherwise, and define the total, $J$, and specific, $j$, angular
momenta within three characteristic radii of the prime halos, with respect
to their CoM.
The correlations between $J$ of baryons and DM are discussed
in \S5.

Both  {$\tilde{J}_{\rm s}$} and {$\tilde{j}_{\rm s}$} end up 
substantially higher in BDM than in PDM model, by a factor of $\sim 5$ --- 
a possible signature of disk angular momentum transfer to the inner 
halo. This is
observed despite that \rts{} in BDM samples a smaller region than in PDM.
Moreover, {$\tilde{J}_{\rm s}$} and {$\tilde{j}_{\rm s}$ in BDM are
nearly constant with time while those in PDM
are variable and show a decline toward $z=0$.
We have tested the radial distribution of $j(M)$, where $M(R)$ is
the DM mass within $R$. At $z=0$, $j(M)\sim M^s$ and $s\sim 0.91$
and $\sim 0.96$, in the PDM and BDM halos, respectively. This
lies within the acceptable range for $s$ in the statistical study 
$j$ distribution of Bullock et al. (2001, see their Fig.~16).

While the values of $J_{\rm vir}$ in both models do not change by
more than a factor of $\sim 2-3$ after the major mergers, 
the orientation of ${\bf\vec J}_{\rm vir}$ changes with the
characteristic timescale of $\sim 1-2$~Gyr by `flip-floping' 
$\sim \pi-2\pi$ angle. These events are not
limited to the major merger epoch. This effect has been already noted
by Porciani, Dekel \& Hoffman (2002)

\subsection{Energy Constraint on Radial Motion in DM Halos}

For the `collisionless' particle motions in
galaxies, the mean free path is much larger than the size of the
system. However, such particle motions are usually limited by energy
and angular momentum considerations, or by their combination. 
We start with the simplest question. How does the energy
consideration limit the particle motions in a halo whose figure
rotation is negligible?

In both models presented here, about $4\times 10^{12}~\msun$ of the
halo have collapsed by $z=0$ and the oscillations have been quickly 
damped
(Fig.~2). In \S4, we show that the virialized part of the halo is
dominated by the DM currents, is clumpy and not fully stratified. 
In view of this, we ask the following question: what fraction of
particles found within $R$ at time $t$ is confined within this radius? 
Specifically, is the
mass within each spherical surface $R$ made up of predominantly
localized (i.e., locally bound) particles or freely streaming
particles across this boundary? 

The motion of
halo particles has corollaries for the disk-halo interaction and
angular momentum exchange and we pursue this elsewhere.  Here we
attempt to develop a number of indicators which {\it quantify} the
dynamical state of the halo during and following the merger epoch.
   
We first estimate the fraction of the bound particles, $\eta(R)$, 
within $R$ as a
function of $z$ in the PDM and BDM prime halos by counting the number
of particles with a total energy below some (negative) value. Because
the halos are non-spherical, this value corresponds to three different
radii along the halo principal axes --- the smallest radius is used
here.  The bound fraction within $R$ is
obtained by dividing the number of bound particles by the total number
of particles within this $R$ at a particular moment of time.

For a comparison, we use the non-singular isothermal sphere. The total 
mass of the NSIS is about $20\%$ higher than \mvir{}
at $z=0$. The NSIS is created close to equilibrium and is allowed to
relax.

We present the radial distributions of the fraction of bound
particles in the halos and in the NSIS, at various $z$. The radii are
normalized by \rvmax{} in all cases for a direct comparison. The
fraction of bound particles in the models increases with time at high
$z$, but stays remarkably constant after the epoch of major mergers,
$z\ltorder 2$ (Fig.~7). At later times, $\eta$ is relatively
flat with $R$, except within \rts. It is about $11\%$ and 
$15\%$ at \rts, $17\%$ and $23\%$ at
\rvmax, and $43\%$ and $57\%$ at \rvir, for PDM and BDM
respectively. Closer to the center, within the NFW cusp region, $\eta$
falls dramatically below $1\%$.

\begin{figure}[ht!!!!!!!!!!!!!!!!!]
\begin{center}
\includegraphics[angle=0,scale=0.43]{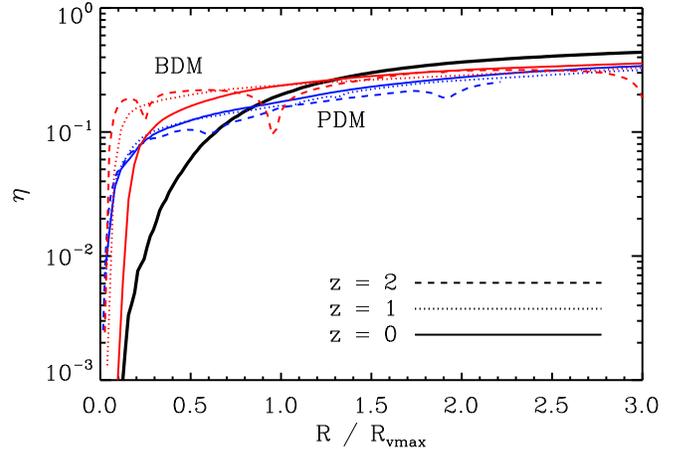}
\end{center}
\caption{Bound fractions of DM, $\eta(R)$, in the prime
  halo of the PDM (blue) and BDM (red) models and in the non-singular
  isothermal sphere, NSIS (thick black). All radii are normalized by 
  \rvmax, for a direct comparison. The lines correspond to
  $z=2$ (dashed), $z=1$ (dot-dashed) and $z=0$ (thin solid). The
  wiggles at higher $z$ are due to the presence of subhalos.  }
\end{figure}

The bound fractions within the PDM and BDM models and the NSIS agree
nicely at large radii.
Within \rvmax, however, the BDM halo is more bound, especially at
$z\gtorder 1$ when its $\rho_{\rm DM}\sim R^{-2}$ because of the
adiabatic contraction. Even the PDM is more bound here than NSIS.
However, closer in, within the cusp region, the
flat core of the BDM is less bound than the NSIS and especially of
the PDM.  Consequently, one
should expect relatively larger radial excursions of the NSIS particles
throughout \rvmax{} compared to the PDM and BDM. Within the flat core, 
the BDM particles will exhibit the larger radial excursions.

Overall, the energy consideration in the halo provides a very weak
constraint on the DM particles motions --- a large fraction of
particles are not bound within a particular halo region but exersize
large radial motions. The next question is whether these motions 
correspond to a coherent behavior of a large number 
of particles. We attempt to answer this question in \S4.

We find that about $50\%$, $20\%$ and $13\%$
of DM particles in PDM and BDM models within \rvir, \rvmax{} and \rts,
respectively, are bound at $z=0$. Fig.~8 shows that these bound 
fractions, $\eta$, are nearly
constant with time within \rvmax{} and \rts, except the fraction within \rts{} 
in the BDM which declines slowly. The bound fractions within \rvir{}
show a secular growth which mirrors that of the \mvir{} itself. 

\begin{figure*}[ht!!!!!!!!!!!!!!!!!!]
\begin{center}
\includegraphics[angle=0,scale=0.3]{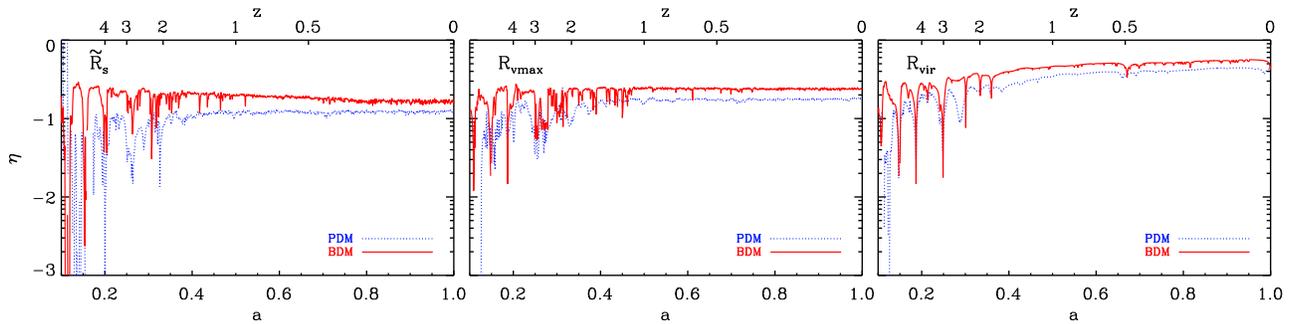}
\end{center}
\caption{Evolution of bound DM fractions, $\eta$, in PDM and BDM models 
within \rts{} (left), \rvmax{} (middle) and \rvir{} (right).}
\end{figure*}

We also confirm that the {\it bound} DM mass within \rts{}
assemble by the end of the merger period 
(e.g., Wechsler et al. 2002), as well as for \rvmax. However, as 
exhibited by Fig.~8, these fractions within \rts{} and \rvmax{} are small
and most of the mass is contributed by the locally unbound particles.  
Therefore, one cannot conclude that the mass accumulation within these
radii terminates with the major mergers.

\subsection{Subhalos Contribution to the Prime Halo Buildup}

While the evolution of the subhalo population is addressed in the
companion Paper~II, here we estimate its contribution
to the buildup of the prime halo mass within \rts, 
\rvmax\ and \rvir. We are not only interested in the contribution 
by the major mergers, i.e., mergers with the mass ratios $\gtorder 1:3$, 
but also smaller ratios of $\gtorder 1:10$ and down to $\sim 10^{-4}$, 
which we (arbitrary) consider as the limit of clumpy accretion. This
corresponds to the subhalo mass of $4\times 10^8~\msun$, which is still well
resolved in our simulations. Any accreted clump below this value is 
considered a part of a smooth accretion.
We also analyze the history of DM 
particles bound within the characteristic radii. 

\begin{figure}[ht!!!!!!!!!!!!!!!!!!]
\begin{center}
\includegraphics[angle=0,scale=0.45]{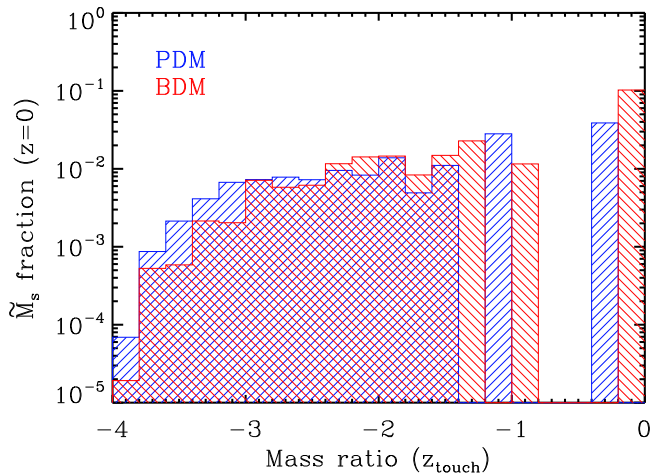}
\end{center}
\caption{Fraction of the subhalo population mass residing within 
\rts\ at $z=0$ in the PDM (blue) and BDM (red) models as a 
function of the merger mass ratio (subhalo-to-prime halo, see text)
calculated at `touching' $z_{\rm touch}$. 
DM accreted with mass ratios less than $10^{-4}$ is accounted for
as a smooth accretion. This value corresponds to a subhalo of
$4\times 10^8~\msun$, well resolved in these simulations. 
}
\end{figure}

We briefly outline our two methods of calculating the merger
mass ratios. The subhalo mass is calculated after determining its tidal 
radius. In the first method, we determine the tidal radii using the 
spherical overdensities of the prime halo and
subhalos crossing \rvir{} of the prime. We associate redshift $z_{\rm touch}$ 
with subhalos entering this radius. The second method is based on the
prime halo and subhalos density isocontours and does not involve
sphericalization. We determine the 3-D isodensity contours and define
the `virial' contour $\rho_{\rm vir}\equiv \Delta(z) \rho_{\rm crit}$ 
as the halo boundary. The subhalo mass is taken as its tidally-truncated
value when it crosses $\rho_{\rm vir}$. In this paper we use the
latter definition of merger mass ratios and compare it with the former
one in Paper~II. 

About $15\%$ and $33\%$ of \mts{} (the DM only mass in PDM and BDM) 
contributed by subhalos is found within \rts{} of the prime
halos at $z=0$, respectively (Fig.~9). That includes 
$4\%$ and $20\%$ from the major mergers, and $11\%$ and $15\%$, 
respectively, from the minor mergers. The rest came from the smooth 
accretion. 
Overall, contribution to \ms{} and \mts{} from mergers, when counted
per decade of the mass ratio, falls off slowly (Fig.~9), from 1:1
to $10^{-3}$. Lower mass ratio mergers contribute little to the central
mass, but the total contribution from smooth (unresolved) accretion
dominates.

\section{Results: Kinematics of DM Buildup in PDM and BDM models}

Hydrostatic equilibrium refers to an internal state when the inward
force of gravity is balanced by the pressure gradient force. The
fluid element must also be at rest, so no large-scale motions are
allowed. The latter condition is not strickly
fulfilled in the halo which forms within the hierarchical merging
scenario. 
The pressure gradient term in collisionless systems is given by the
gradient of a stress tensor, whose off-diagonal terms are comparable
in magnitude to the diagonal ones and are responsible for the
non-conservation of circulation. Therefore, in collisionless systems
these terms drive evolution on much shorter dynamical timescales than
in fluids, where the off-diagonal terms are small because they
represent viscous forces (Christodoulou et al. 1995).

The drive of the DM halos towards the virial equilibrium is
accompanied by the violent relaxation (Lynden-Bell 1967). However the
efficiency of violent relaxation depends dramatically on the presence
of residual currents in the halo, in liu of the quickly decaying
central potential oscillations. Such currents are in fact favored in 
our simulations
because of the presence of large-scale filaments, as they are the
natural outcome of cosmological initial conditions. The main
and secondary filaments (e.g., Fig.~1) form at high $z$.
We therefore study the kinematics of these currents and
analyze the mixing they introduce in the halo.

The filaments consist of subhalos and a less clumpy material that
penetrate \rvir, preserve their identity for at least
one crossing time (depending on their impact parameter with the main
halo) and display dispersion velocities which are substantially lower
than in the surrounding halo.
In fact, we observe that the `cold' filament-driven influx is
de-focused after passing the pericenter of its motion. As its
constituents move out, their slowdown in tandem with the tidal
disruption lead to the formation of shells that persist for a long
time. Shells that form later in time have larger outflow velocities
and can cross shells that formed earlier (Paper~II). For the
survival of these shells it is important that they form after subhalos
pass the pericenters of their orbits.

When discussing the main halo buildup, we distinguish between the
first-time entering material and the recycled material. The latter 
consists of shells, tidal tails and captured subhalos.  In the
following we continue to abbreviate the spherical components with $R$
and $v_{\rm R}$ (i.e., particle velocity along $R$), and cylindrical
components with $r$ and $v_\phi$ (the radius and azimuthal velocity in
an arbitrary $xy$-plane whose origin lies at the DM CoM). In 
order to specify the kinematics of the buildup, we
analyze the $R-v_{\rm R}$ (in \S4.1) and $r-v_\phi$ (in \S4.2) diagrams
within the inner 300~kpc and zoom into the innermost 20~kpc when
focusing on the disk buildup.

\subsection{Halo Evolution in $R-v_{\rm R}$ Plane}

We now follow the evolution of DM particles in the $R-v_{\rm R}$ plane. This
reveals the intricacies of halo relaxation beyond virialization --- to
what extent it `forgets' the initial conditions and how the
substructure is gradually erased. Most importantly, it allows the
comparison of the DM kinematics.
\begin{figure*}[ht!!!!!!!!!!!!!!!!!!]
\begin{center}
\includegraphics[angle=0,scale=0.38]{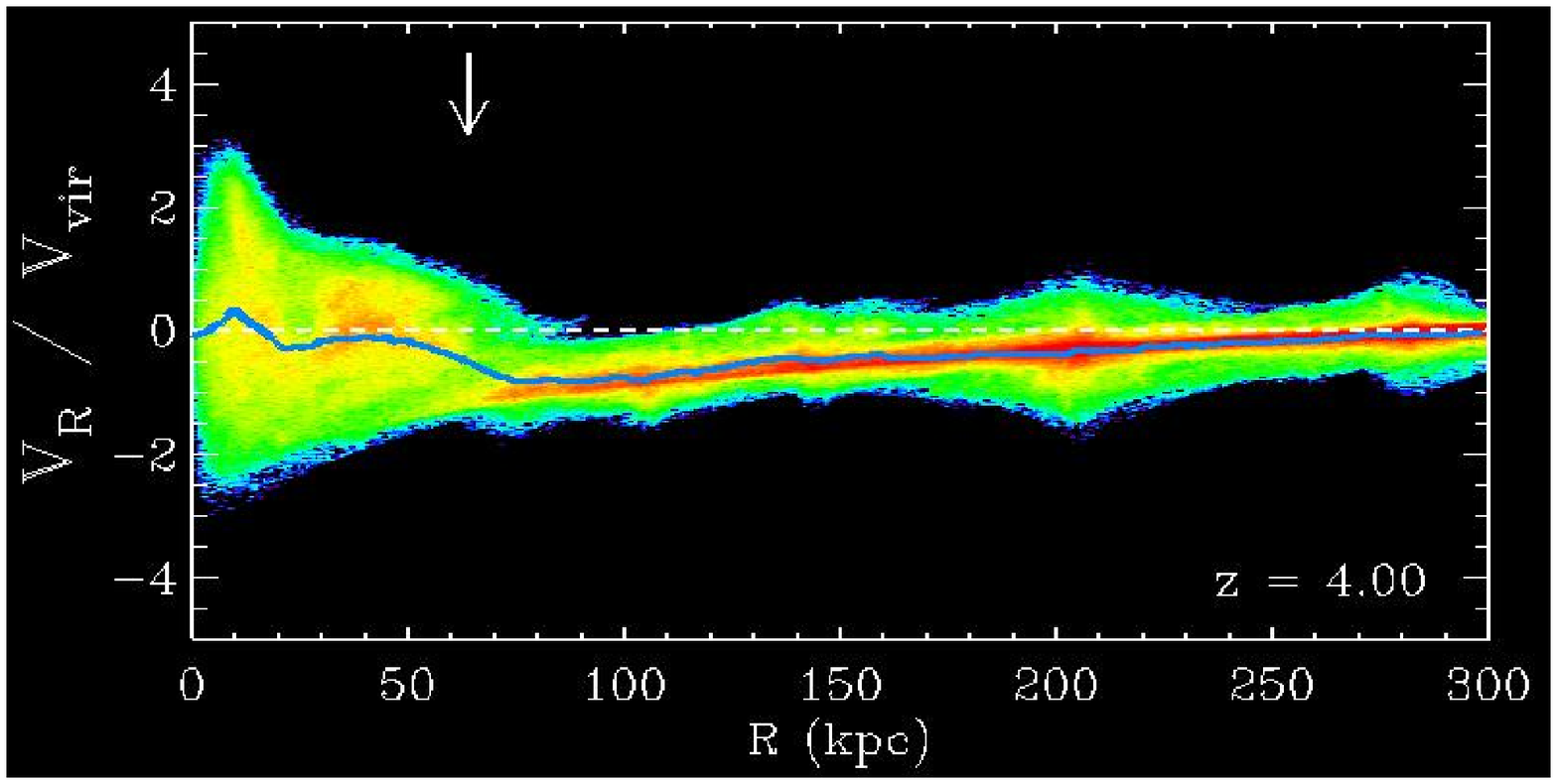}
\includegraphics[angle=0,scale=0.38]{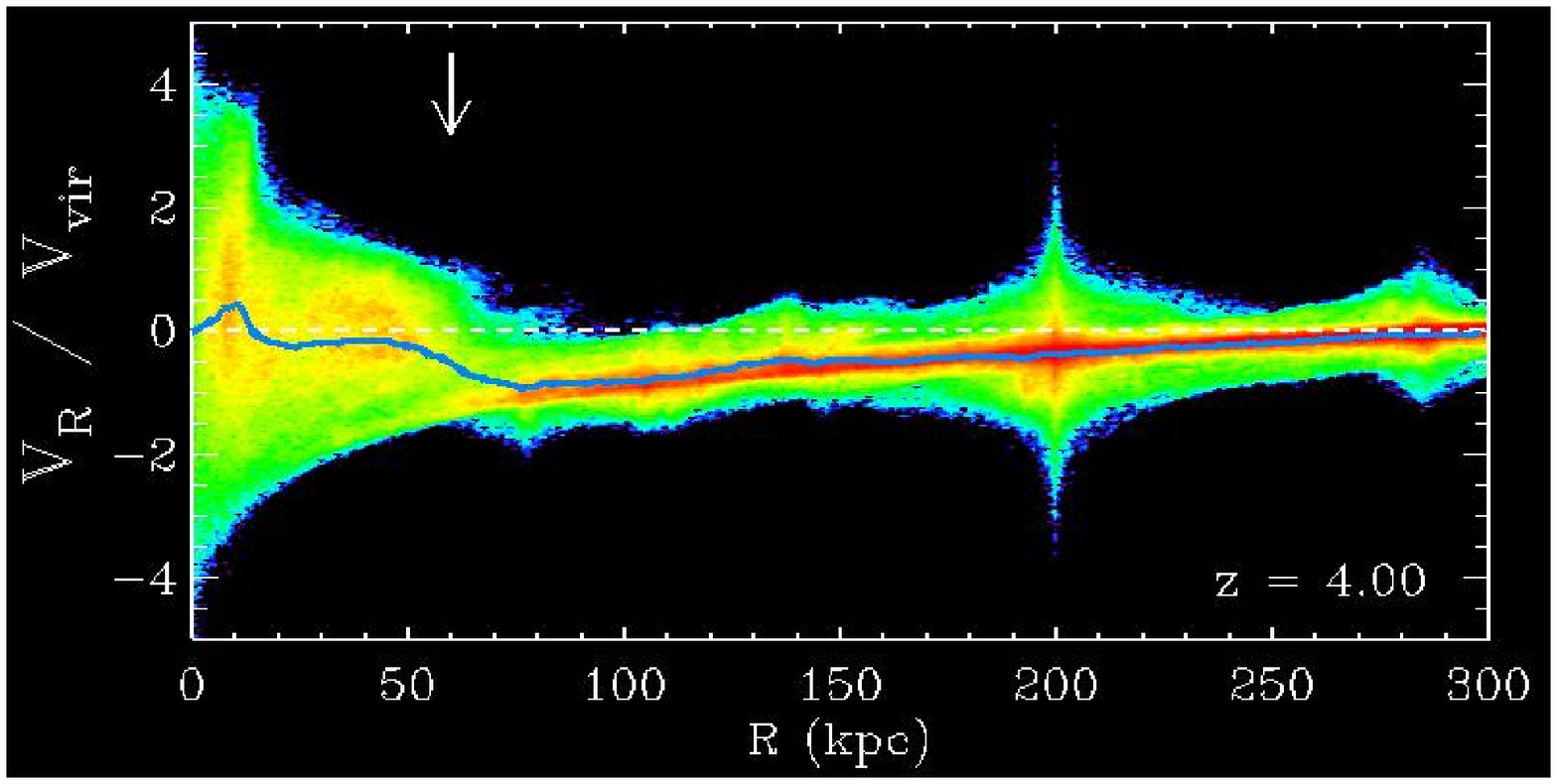}
\includegraphics[angle=0,scale=0.38]{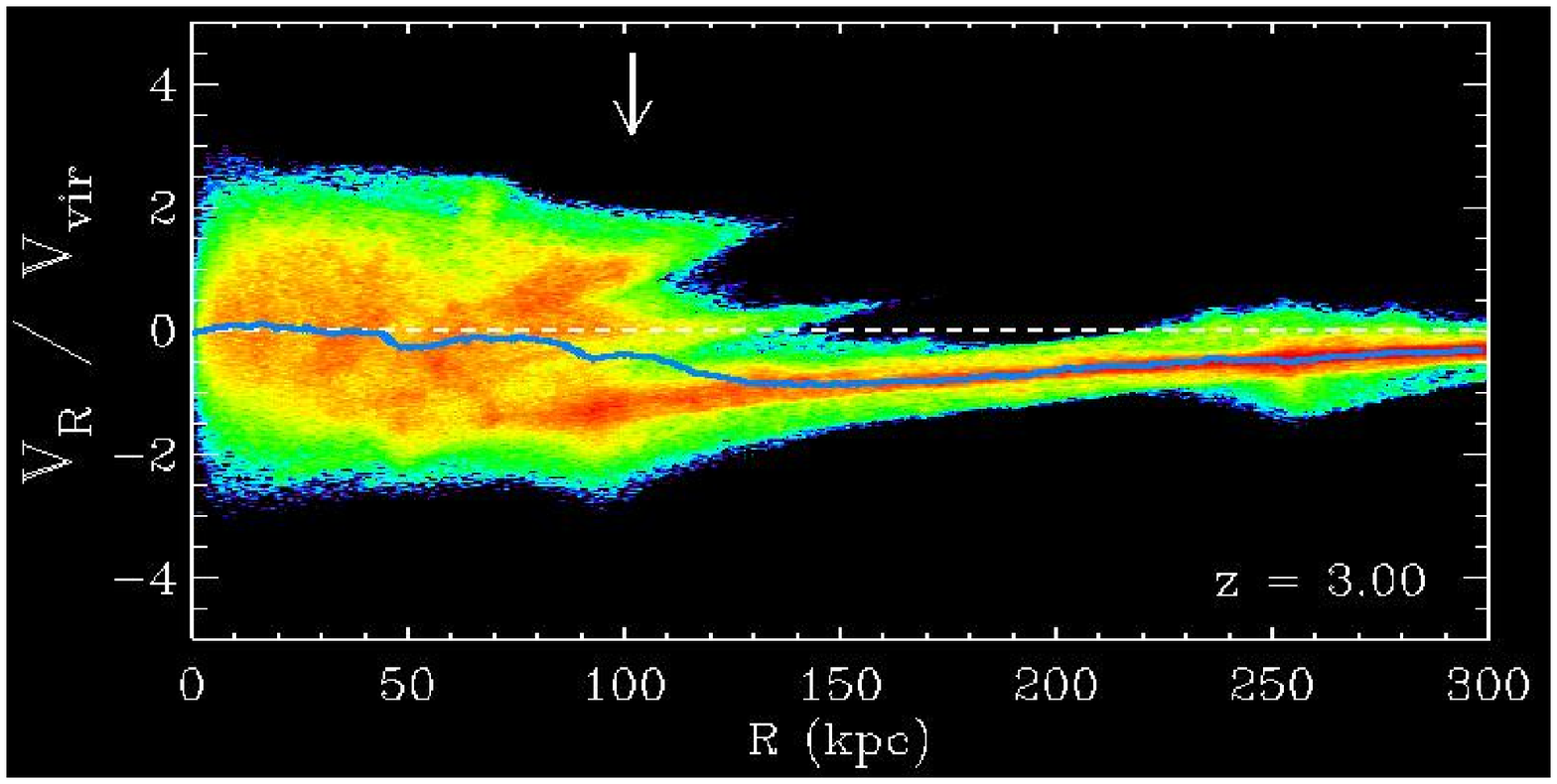}
\includegraphics[angle=0,scale=0.38]{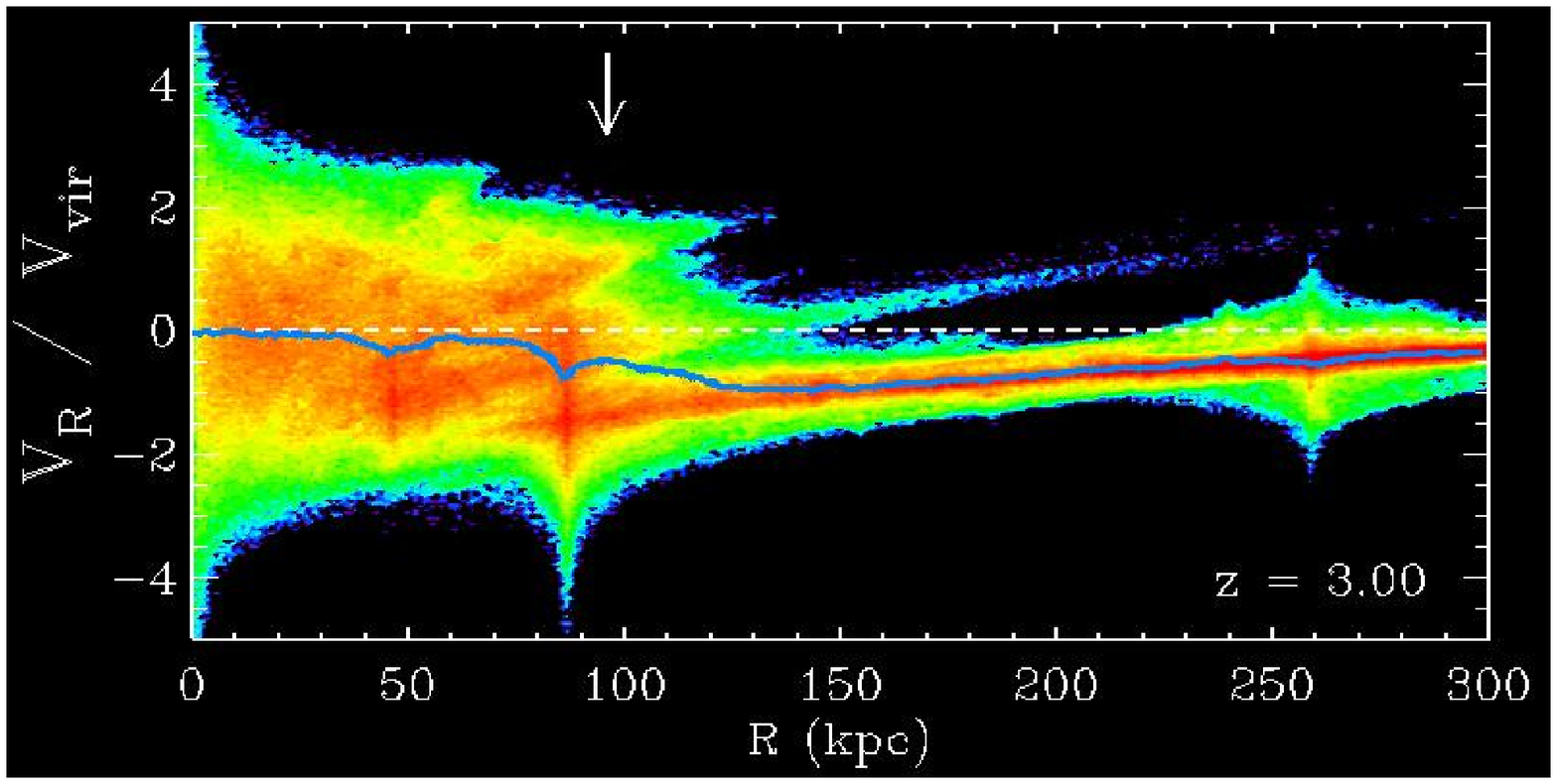}
\includegraphics[angle=0,scale=0.38]{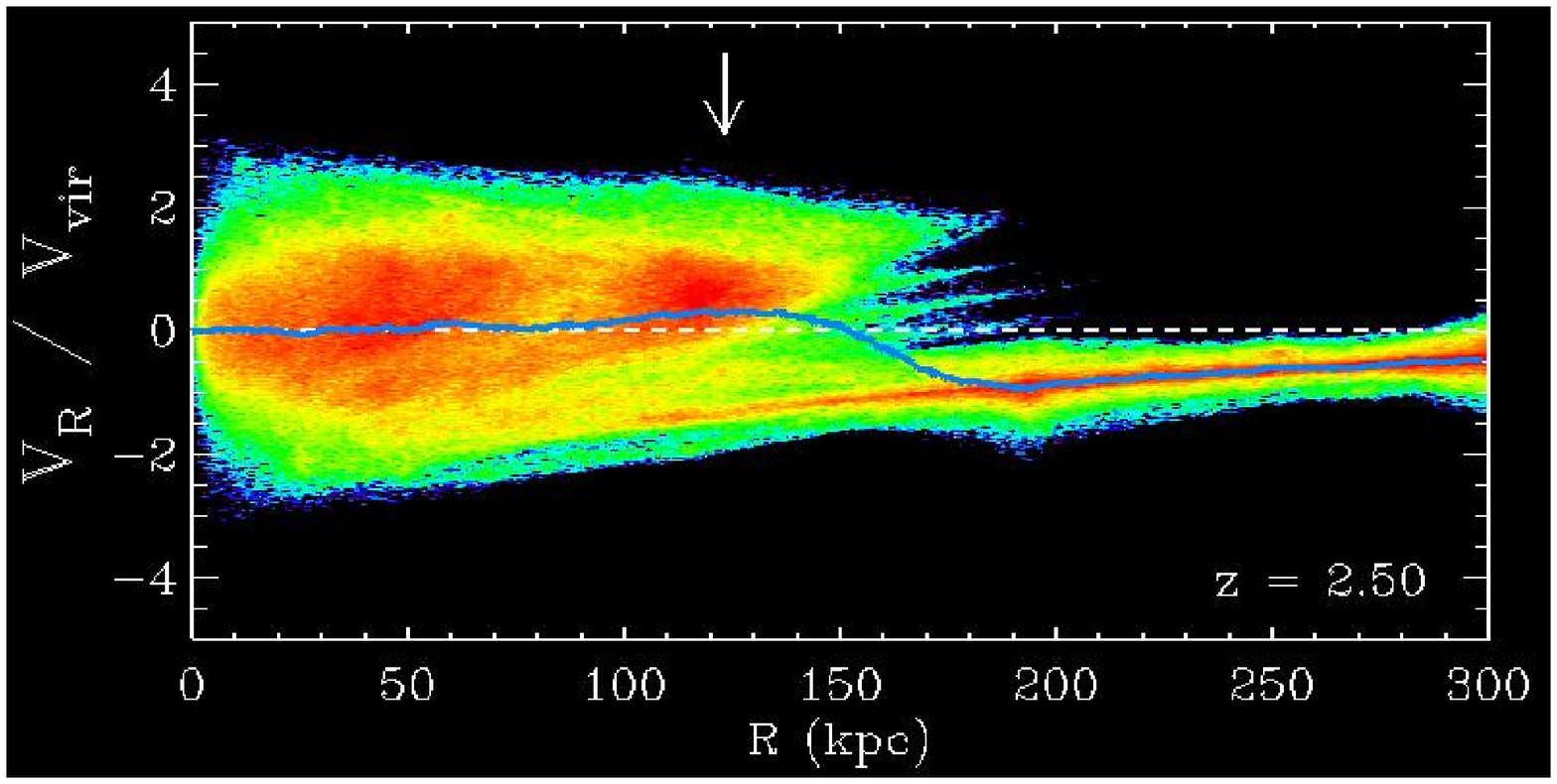}
\includegraphics[angle=0,scale=0.38]{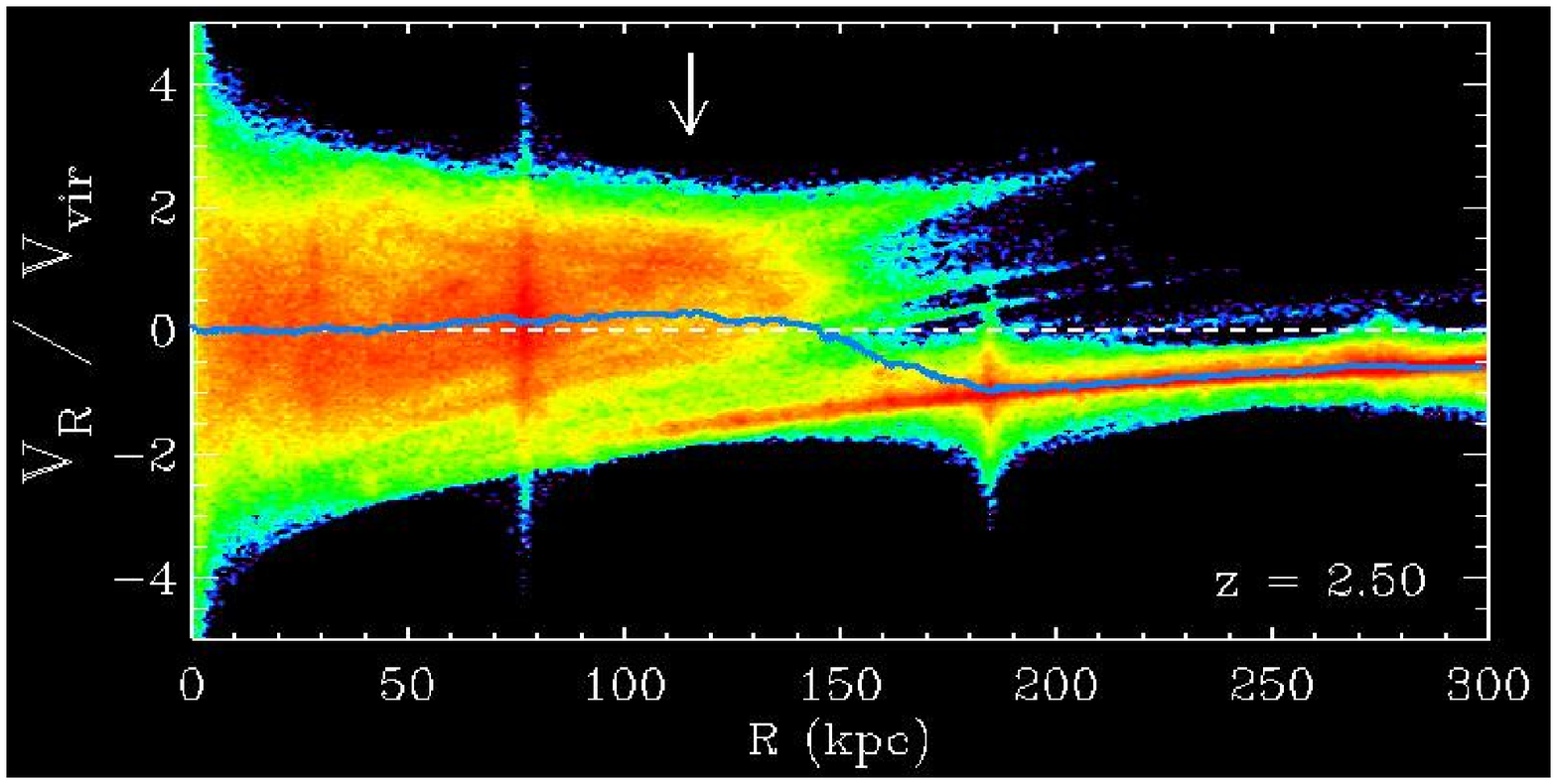}
\includegraphics[angle=0,scale=0.38]{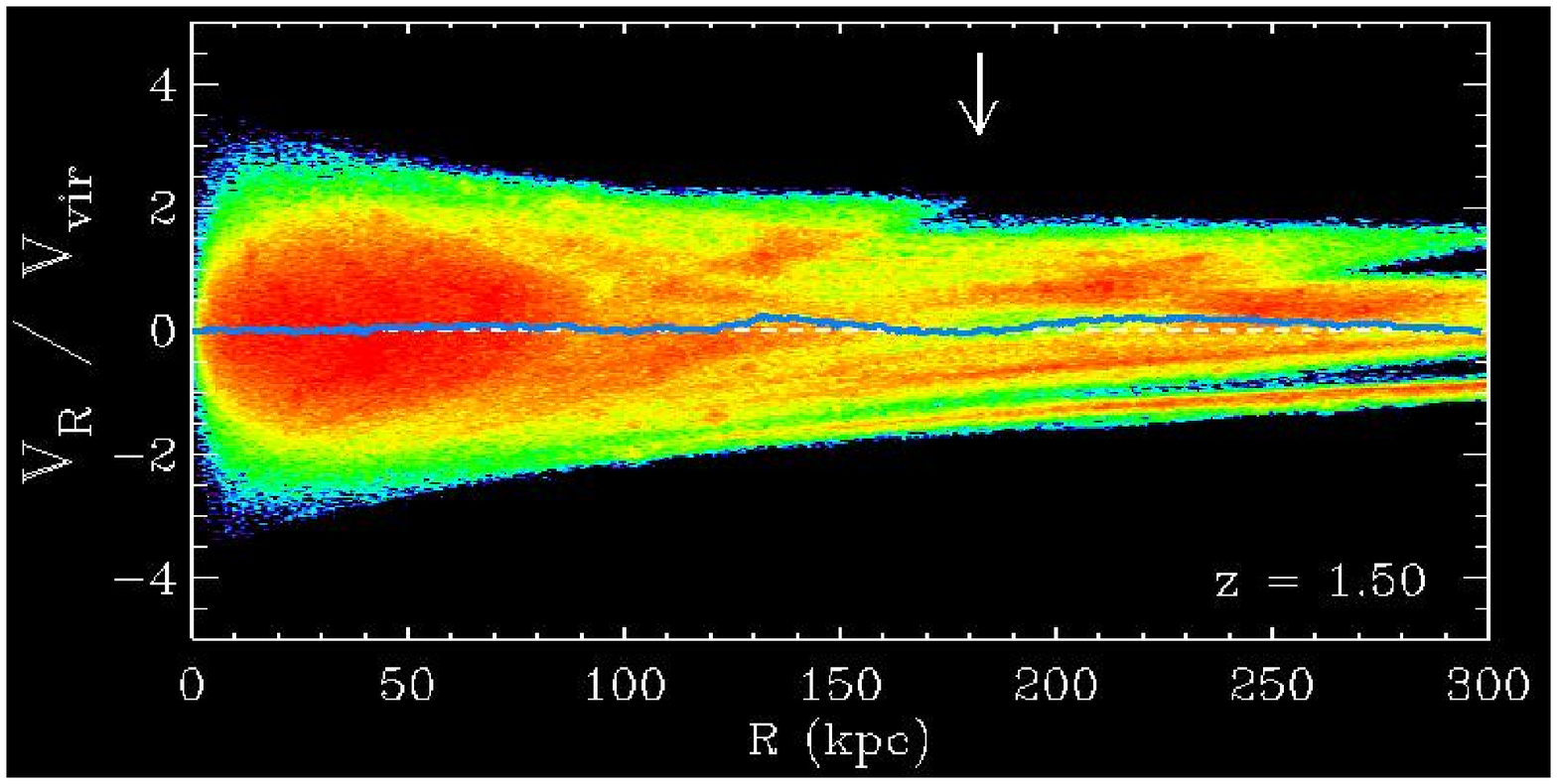}
\includegraphics[angle=0,scale=0.38]{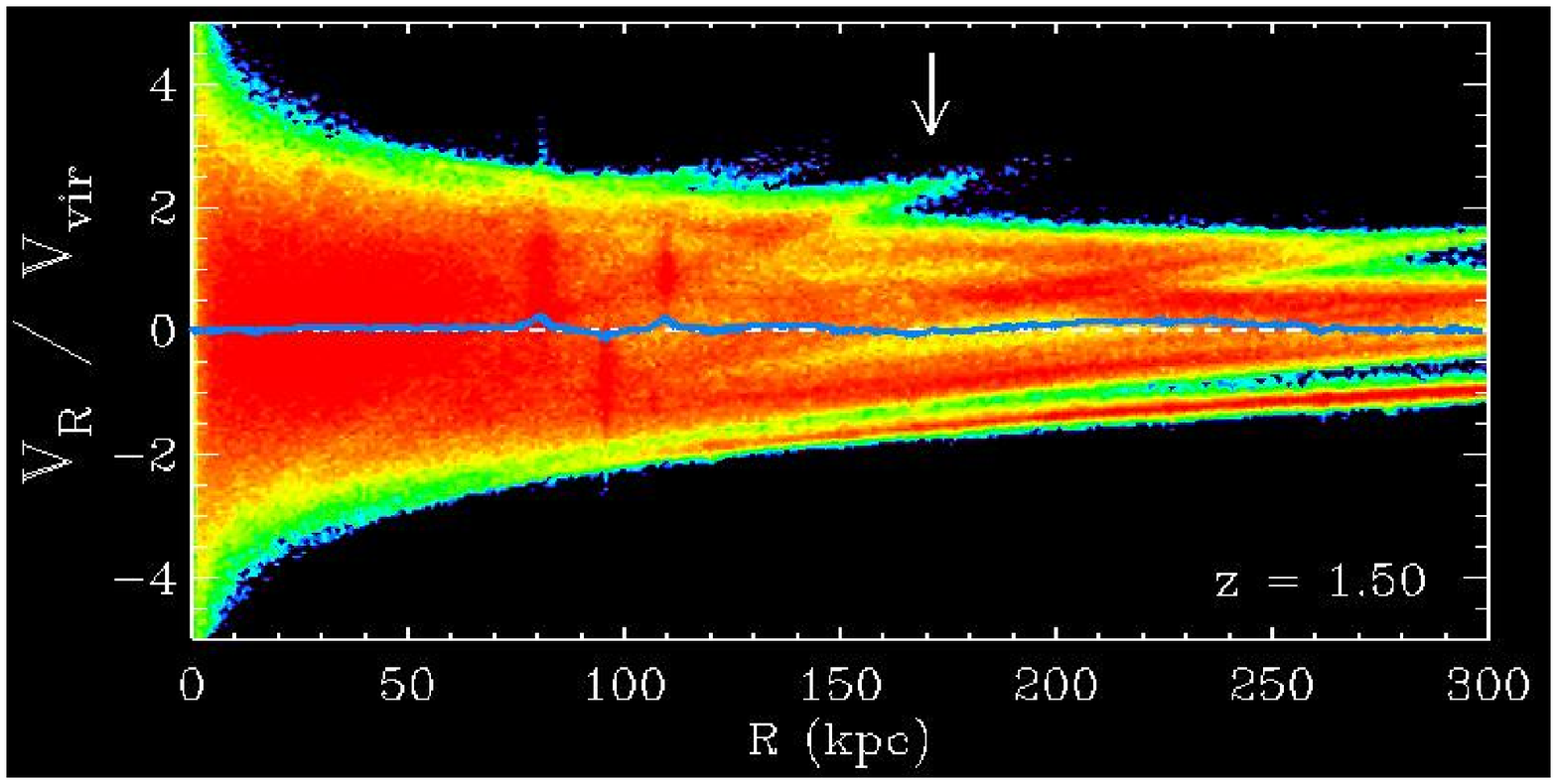}
\includegraphics[angle=0,scale=0.38]{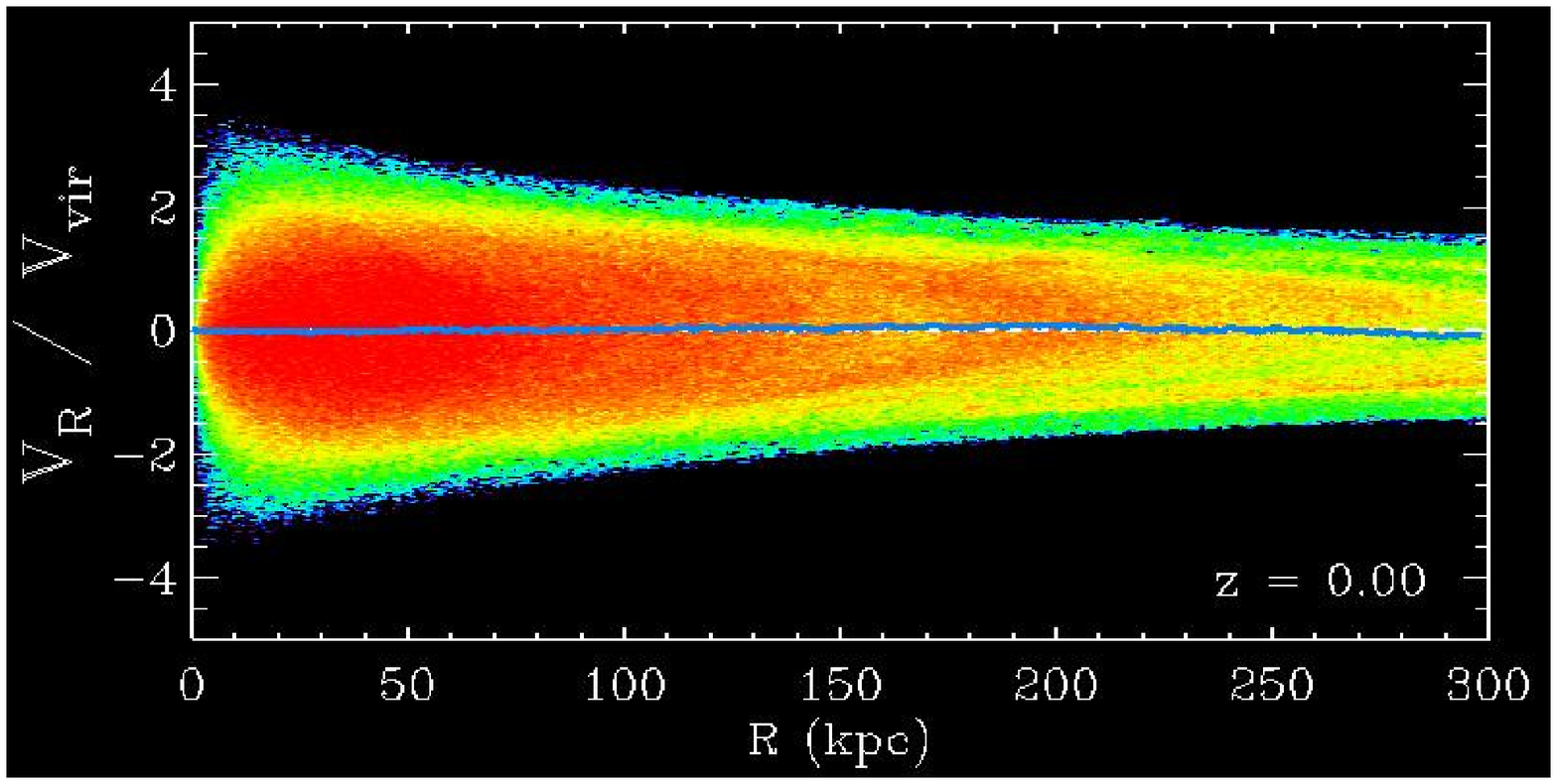}
\includegraphics[angle=0,scale=0.38]{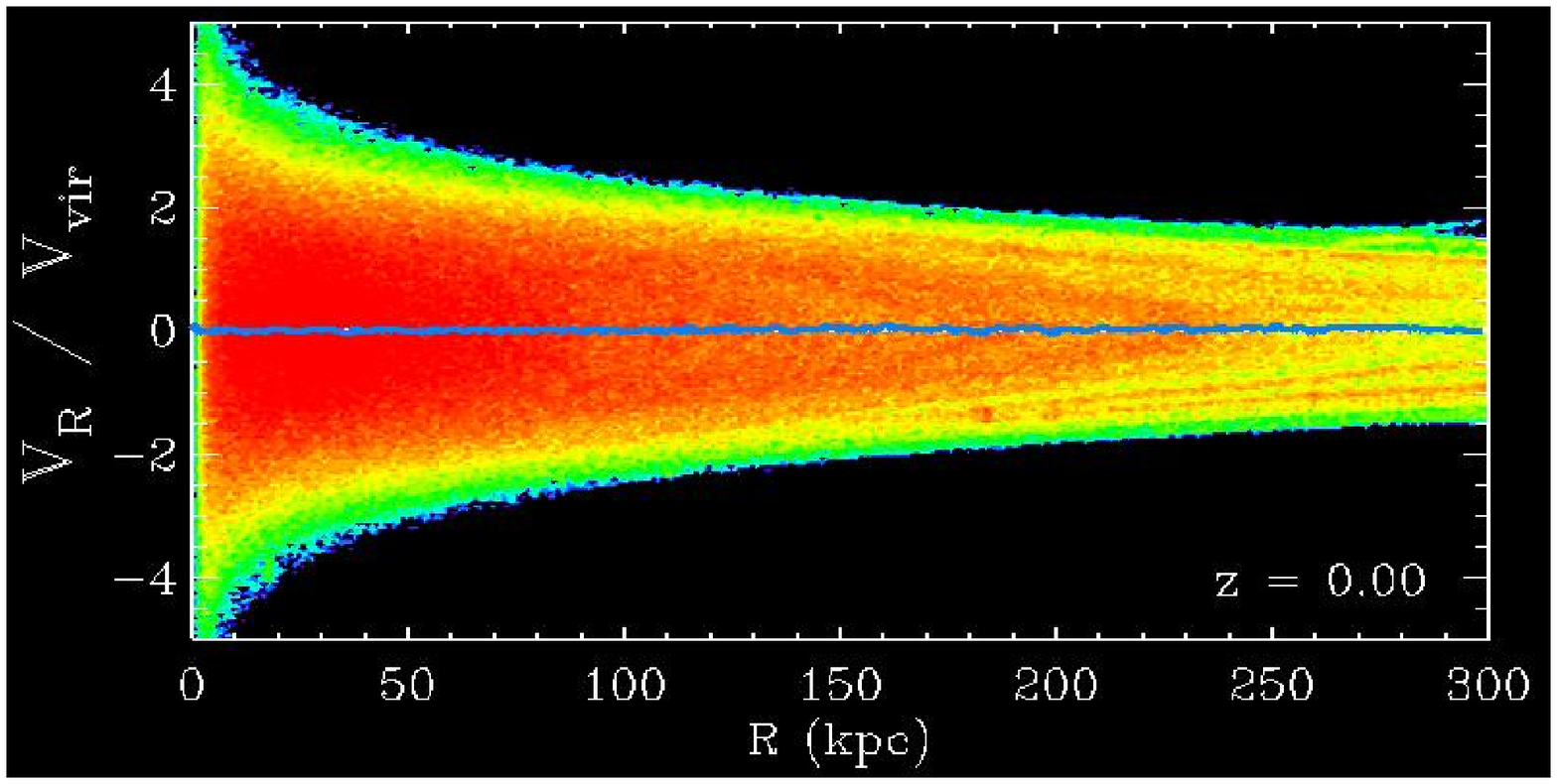}
\end{center}
\caption{Evolution of the DM halo in the PDM (left) and BDM (right)
  models shown in the $R-v_{\rm R}$ plane at following times: 
  $z=4$ and 3 (the major merger epoch
  and the appearance of `fingers,' and the subsequent minor mergers),
  and $z=2.5, 1,5$ and 0 (the appearance of the shell structure inside
  and outside of the main halo). The colors correspond to the DM
  particle density on the $R-v_{\rm R}$ surface. The vertical arrow shows
  \rvir, the dashed white line --- $v_{\rm R}=0$, and blue line --- the
  average $v_R$ at each $R$. The velocity axis is normalized by $v_{\rm vir}$
  --- the circular velocity at \rvir. The associated Animation displays the
  evolution in this plane from $z=25$ to $z=0$.  }
\end{figure*}

The radial velocity field of the DM is dominated by an outflow
initially (i.e., cosmological expansion), which slows down from
inside-out, reflecting the gradual increase in the turnover
radius. Within \rvir, the distribution of $v_{\rm R}$'s is initially slightly
asymmetric with respect to the $v_{\rm R}=0$ line (Fig.~10 and the Animation
Sequence), then gradually decreasing its bias by $z\sim 6$, with the 
negative $v_{\rm R}$ asymmetry remaining only outside \rvir. The growth of
substructure is clearly visible as local (vertical) distortions in the
velocity fields. As the subhalos themselves virialize, these vertical
distortions provide information about their binding energies and
mutual interactions and mergers. The symmetry of each subhalo peak
indicates whether they continue to grow --- the more symmetric is the
peak with respect to their local CoM, the smaller is their growth rate at
this moment.

The epoch of major mergers can be easily followed in Fig.~10. By
$z\sim 4$, the $v_{\rm R}$ field is symmetric (with respect to $v=0$ line)
only within the central 60--70~kpc, and even this is a very
approximate statement because the substructure is far from being
erased there. The main halo buildup can be closely followed --- the
steady stream of virialized subhalos and the smooth (i.e., below our
self-imposed limit) accretion inflow can be clearly distinguished. Outside the
growing halo, within some distance, the prevailing velocities are 
negative, both in smooth and clumpy accretion,
with only more massive subhalos contributing to the positive $v_{\rm R}$
(due to the internal velocity dispersions).  The major mergers appear
as negative $v_{\rm R}$ moving features, then switching to a positive $v_{\rm R}$
at pericenters, producing strong asymmetries.

The destruction of subhalos in tidal interactions and mergers is
visible when the vertical velocity feature (spike) of each subhalo inclines
clockwise. This corresponds to the formation of
tidal tails whose $R-v_{\rm R}$ correlations in particle number density are
naturally reproduced and are visible as `fingers. '

Comparison between different frames in Fig.~10 displays the diverging
evolution of the DM in the simulations. The early evolution is very
similar in both models, but after $z\sim 4$, both the main halo and
the penetrated subhalos increasingly differ. The BDM simulation
achieves higher central velocities because of the presence of baryons
which drag the DM inwards in an adiabatic contraction. The tidal
disruption of subhalos is associated with much more pronounced tidal
tails in the BDM, i.e., inclined `fingers.' The fingers appear
longer by a factor of 2 in the BDM.

At higher $z$, the influx across \rvir{} dominates the buildup, while
at lower $z$'s the trapped subhalos and smooth accretion are
recycled. Some of the subhalos survive a few orbits (vertical fingers),
while other are tidally disrupted (inclined fingers). 
The recycled part of the accreted material 
has a longer `memory at later times --- the individual shells
survive longer and are less mixed.

The BDM halo becomes substantially more concentrated than the prime
halos in the PDM. Note the almost double spread along the $v_{\rm R}$-axis in
the former. Moreover, the (inner) shape of the denser (in color)
region is smashed against the $v_{\rm R}$-axis in the BDM and has
a convex shape in the latter. This difference is explained in terms of
the shape of the gravitational potential within the central region of the
prime halo.

By $z\sim 3.5$, the radial velocities are symmetric with respect to
$v_{\rm R}=0$ within about the central 100~kpc. This region increases to $\sim
200$~kpc by $z\sim 2.5$.  By $z\sim 2$, the correlation fingers extend
well beyond the halo radius, to $R\sim 300$~kpc, beyond 400~kpc at
$z\sim 1.7$, beyond 600~kpc at $z\sim 1.4$, and beyond 800~kpc at
$z\sim 1$. The outflowing material crosses the 1~Mpc radius by $z\sim
0.8$. The width of the inflowing stream declines with time, while that
of the rebounding material increases.  The shell structure is easily
distinguished in the outflowing material and can be traced across the
$v_{\rm R}=0$ line. Material can be followed in its circulation along these
shells up to about 800~kpc at $z\sim 0$, well outside \rvir.

\begin{figure}[ht!!!!!!!!!!!!!]
\begin{center}
\includegraphics[angle=0,scale=0.45]{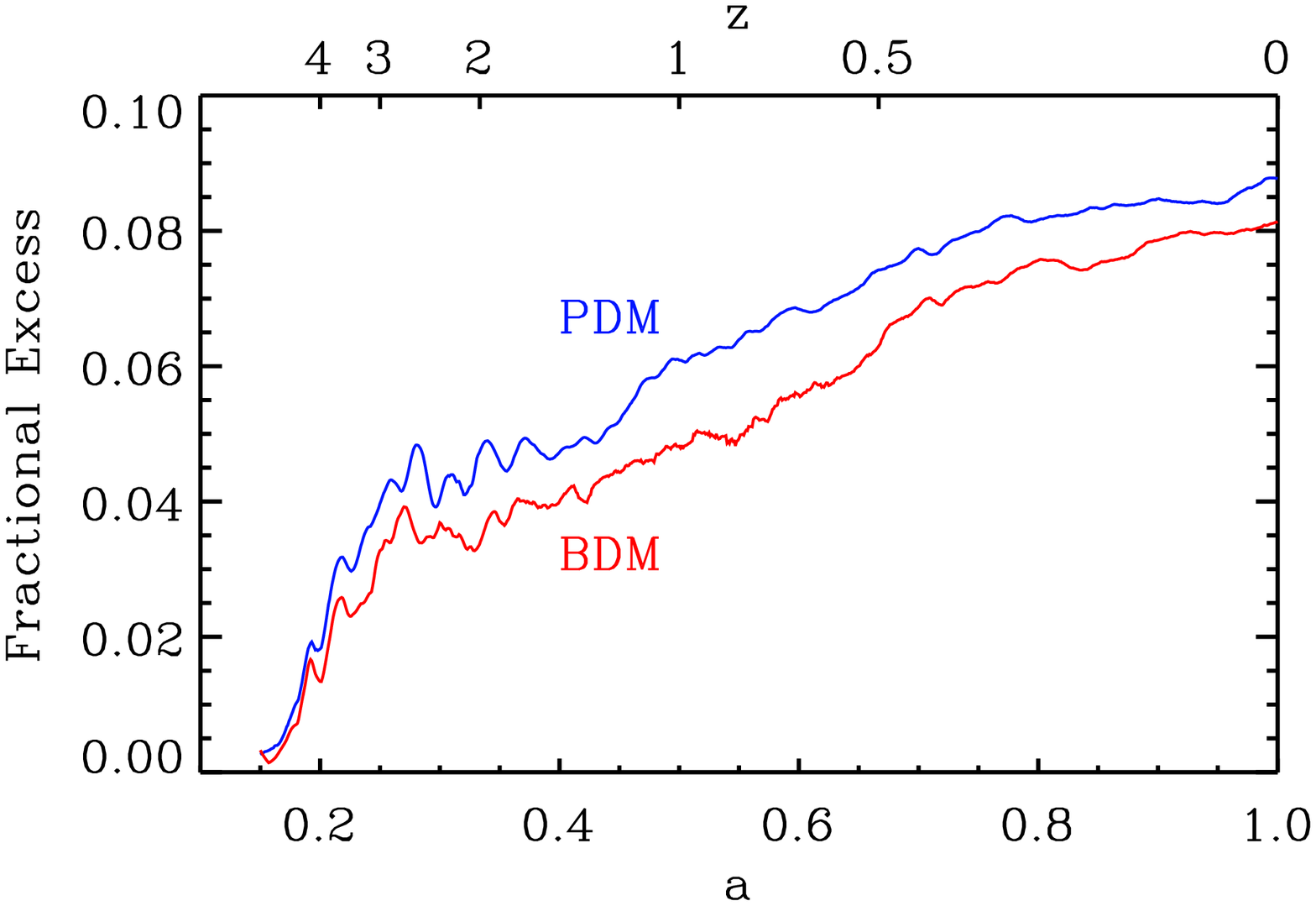}
\end{center}
\caption{Evolution of the {\it excess} fraction of DM mass associated
  with substructure, i.e., with subhalos, tidal tails and streamers,
  in PDM and BDM models.  This fraction is calculated by normalizing
  the residual mass within \rvir\ (e.g., Fig.~10) by \mvir.  }
\end{figure}

\begin{figure*}[ht!!!!!!!!!!!!!]
\begin{center}
\includegraphics[angle=0,scale=0.399]{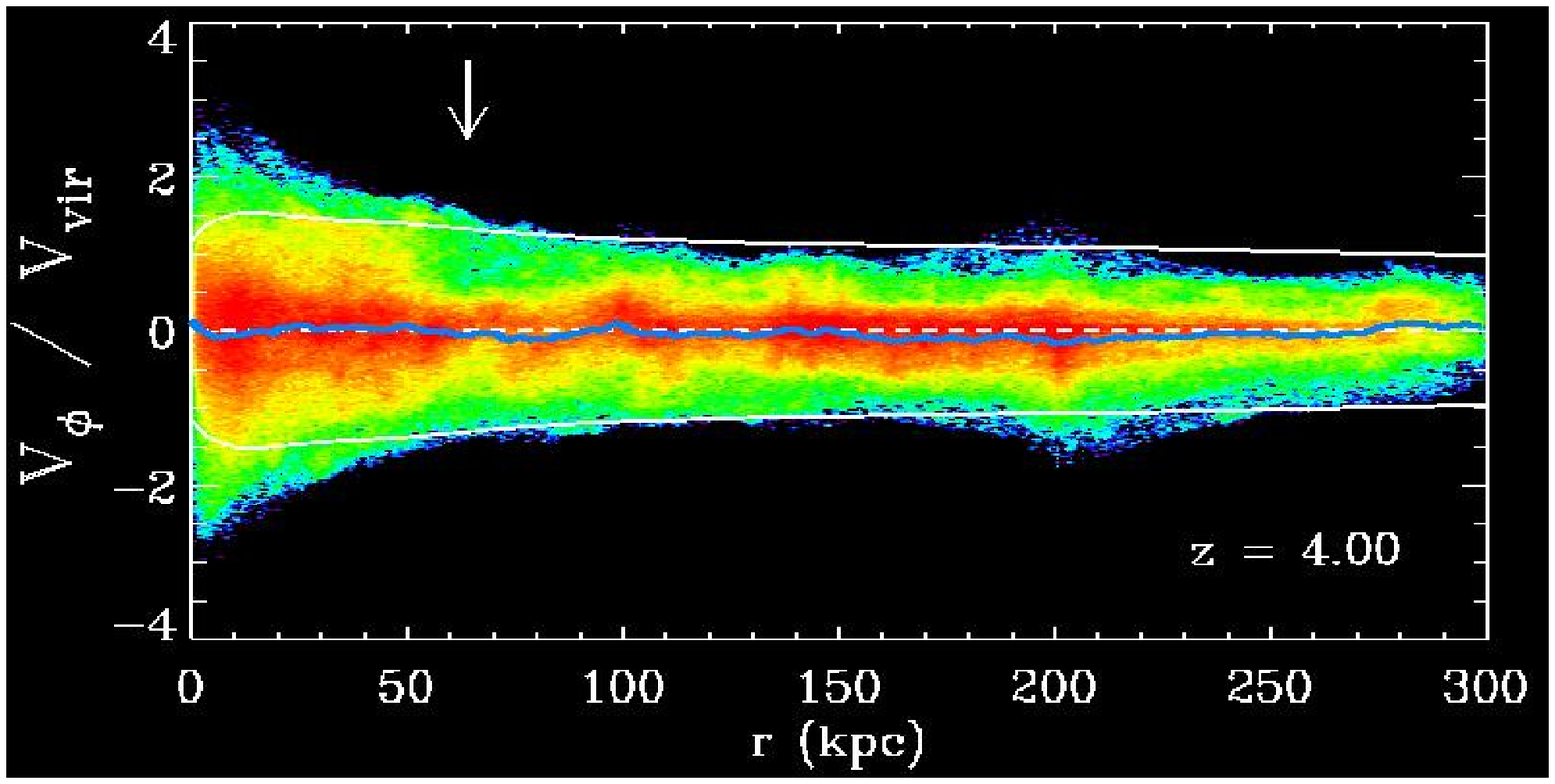}
\includegraphics[angle=0,scale=0.399]{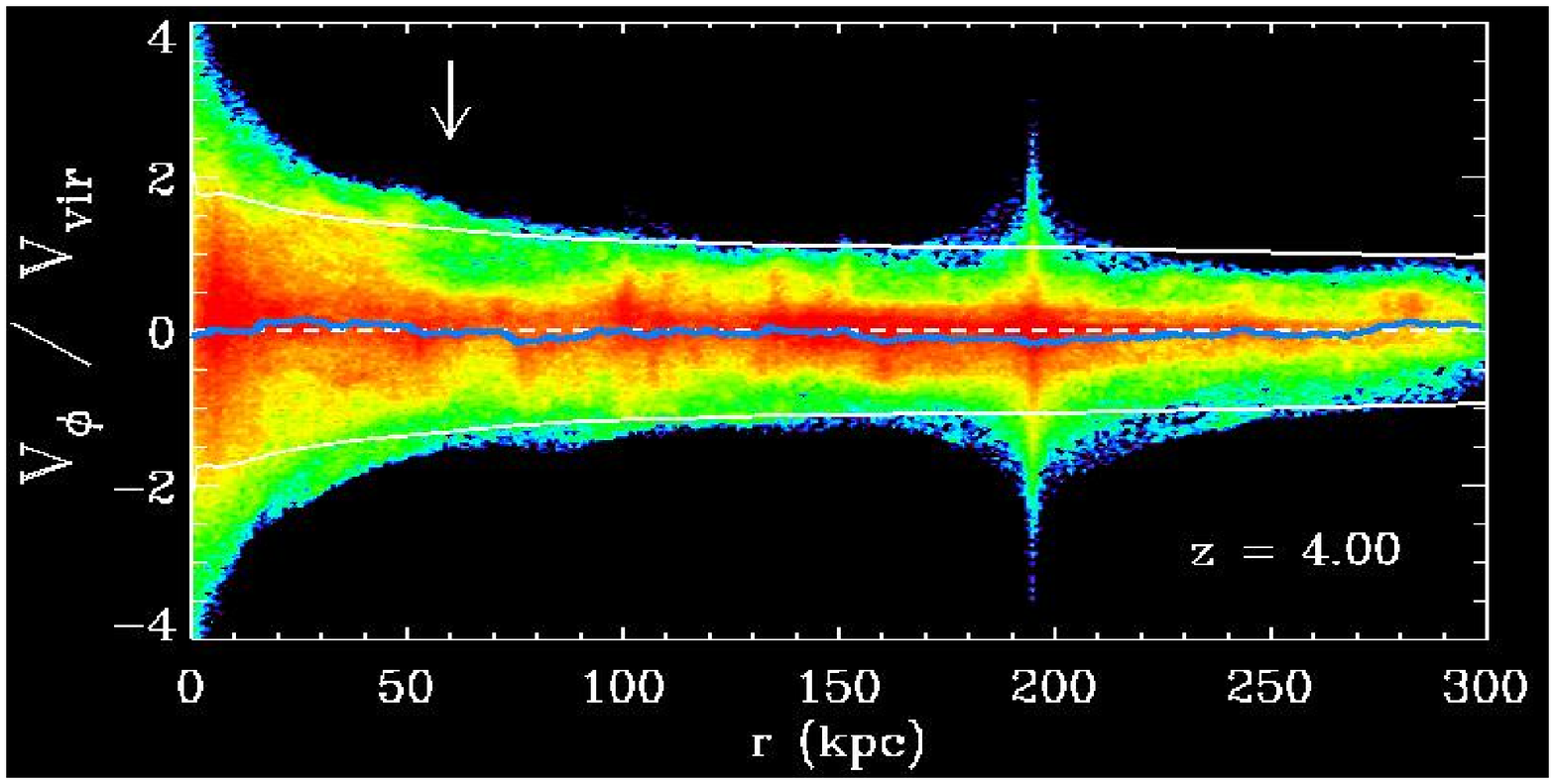}
\includegraphics[angle=0,scale=0.399]{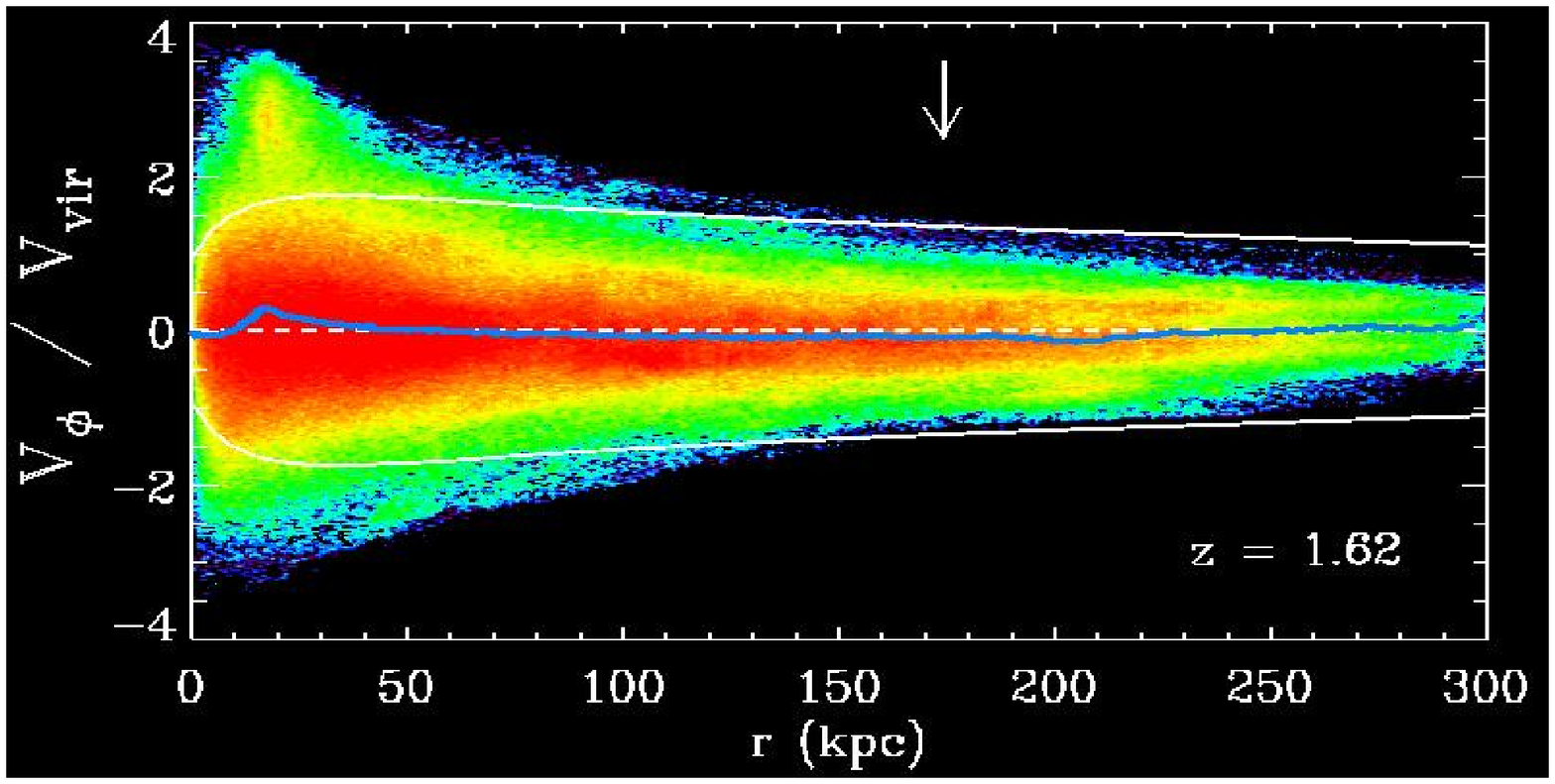}
\includegraphics[angle=0,scale=0.399]{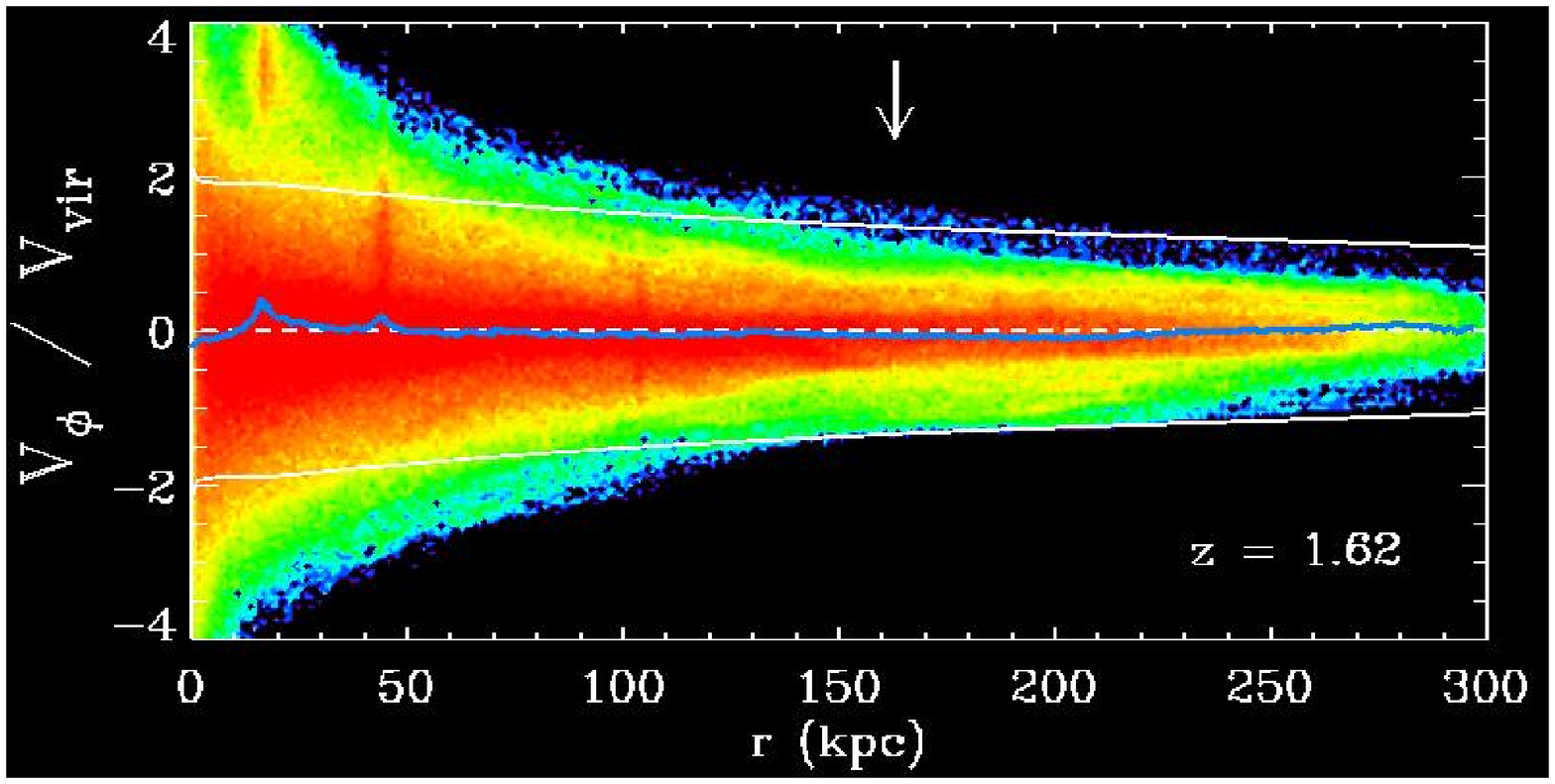}
\end{center}
\caption{Example of evolution of the DM halo in PDM (left) and BDM
  (right) models, shown in the $r-v_\phi$ diagram at following times:
  $z=4$ (the major merger epoch), and $z=1.625$ (minor merger
  event). The colors correspond to the DM particle density on the
  $r-v_\phi$ surface. The white solid lines display the circular
  velocities in the halo as a function $r$ at each $z$. The blue 
  line shows the average $v_\phi$ at $r$. The vertical arrows show the 
  position of \rvir. The velocity axis is normalized by $v_{\rm vir}$
  --- the circular velocity at \rvir. 
}
\end{figure*}

The `cold' stream of incoming subhalos and smooth accretion become
less dominant by $z\sim 2$ and the $R-v_{\rm R}$ diagram is dominated by the
shell structure, but the stream is notiseable even at $z=0$ where it
is the main source of the remaining $v_{\rm R}$ asymmetry within the main
halo. Moreover, the $R-v_{\rm R}$ diagram reveals the absence of large
virial velocities in the central few kpc, except at a few limited
moments when subhalos are passing (or merging) through the
center. This is only a characteristic of the pure DM simulations,
with the baryonic simulations showing large velocities near the center. 
As stated above, this
difference is caused by a much larger central mass concentration in
the latter models.

Lastly, for $z \ltorder 1.5$, three types of motion can be detected in
the $R-v_{\rm R}$ plane. The first one corresponds to the clumpy and smooth
accretion onto the main halo. This flow dominates the bottom parts
of these diagrams at negative velocities.  The top part of the
diagrams, at positive velocities, represents the outflows of the
material processed by the central halo. Initially, it is
dominated by fingers of tidally disrupted subhalos, and at later times
it is dominated by the shells of the ouflowing material. 
At progressively lower redshifts those flow out faster because they
originate at larger distances from the halo and their infalling
velocities increase as well. This causes finger and shell crossings,
observable in the diagrams.  The mid part of
the diagrams is populated by circulating flows which do not escape
\rvir, but show the same finger and shell structure.

As we have seen, the halo while virialized is far from being
relaxed. We distinguish three types of substructure corresponding to
three consequtive levels of relaxation: the subhalos, the subhalos
with tidal tails, and streamers. The subhalos are bound within their
tidal radii, while those with tails are clearly found in the process
of being tidally disrupted.  The streamers are recognizable patterns
in the $R-v_{\rm R}$ plane which are not bound but show correlation in this
plane. They are expected to be mixed with the background in a few
crossing times.

To quantify the degree of relaxation in the main halo, we have
subtracted the smoothed version of the halo in the $R-v_{\rm R}$ plane from
the original frame. The smoothing kernel is about 15~kpc along the
$R$-axis and $75~{\rm km~s^{-1}}$ along the $v_{\rm R}$-axis, and the
smoothing procedure is mass-conserving. By subtracting
the smoothed image from the original one we get the residual map.
We integrate the mass associated with the
residuals to obtain the {\it excess} mass (i.e., over the smoothed
halo) associated with the substructure within \rvir. 
This procedure allows us to
estimate the contribution of the excess DM mass fraction associated 
with density enhancements (i.e., subhalos, tidal tails and streamers)
above some smoothed reference density which is
time-adjusted. Fig.~11 shows the
evolution of this fractional excess mass.   We observe two trends: 
the excess mass fraction in the
substructure becomes more prominent with time and it is marginally
larger in the PDM model. In Paper~II we show that the small excess
associated with the PDM model originates outside the inner $\sim 100$~kpc
virialized region. 

Despite some differences in the definition of subhalos,
we compare our results with those of the Aquarius simulation (Springel
et al. 2008). The latter arrive at $\sim 11\%$ contribution from subhalo
mass within the prime halo \rvir{} at $z=0$. Fig.~11 shows about $8\%-9\%$
contribution in our PDM and BDM models at the same time.

\subsection{Halo Evolution in $r-v_\phi$ Plane}

\begin{figure*}[ht!!!!!!!!!!!!!!!!!!]
\begin{center}
\includegraphics[angle=0,scale=0.408]{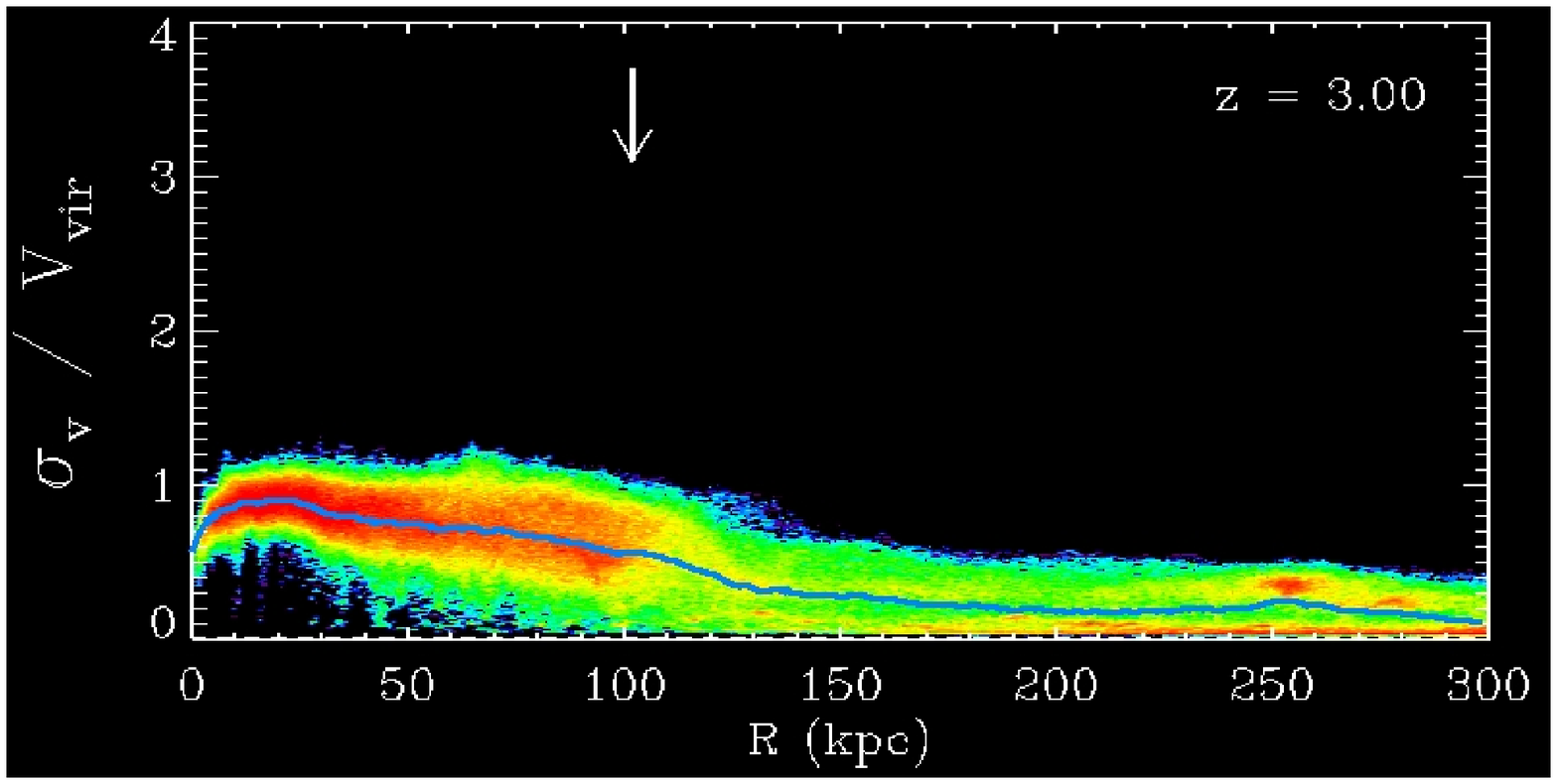}
\includegraphics[angle=0,scale=0.408]{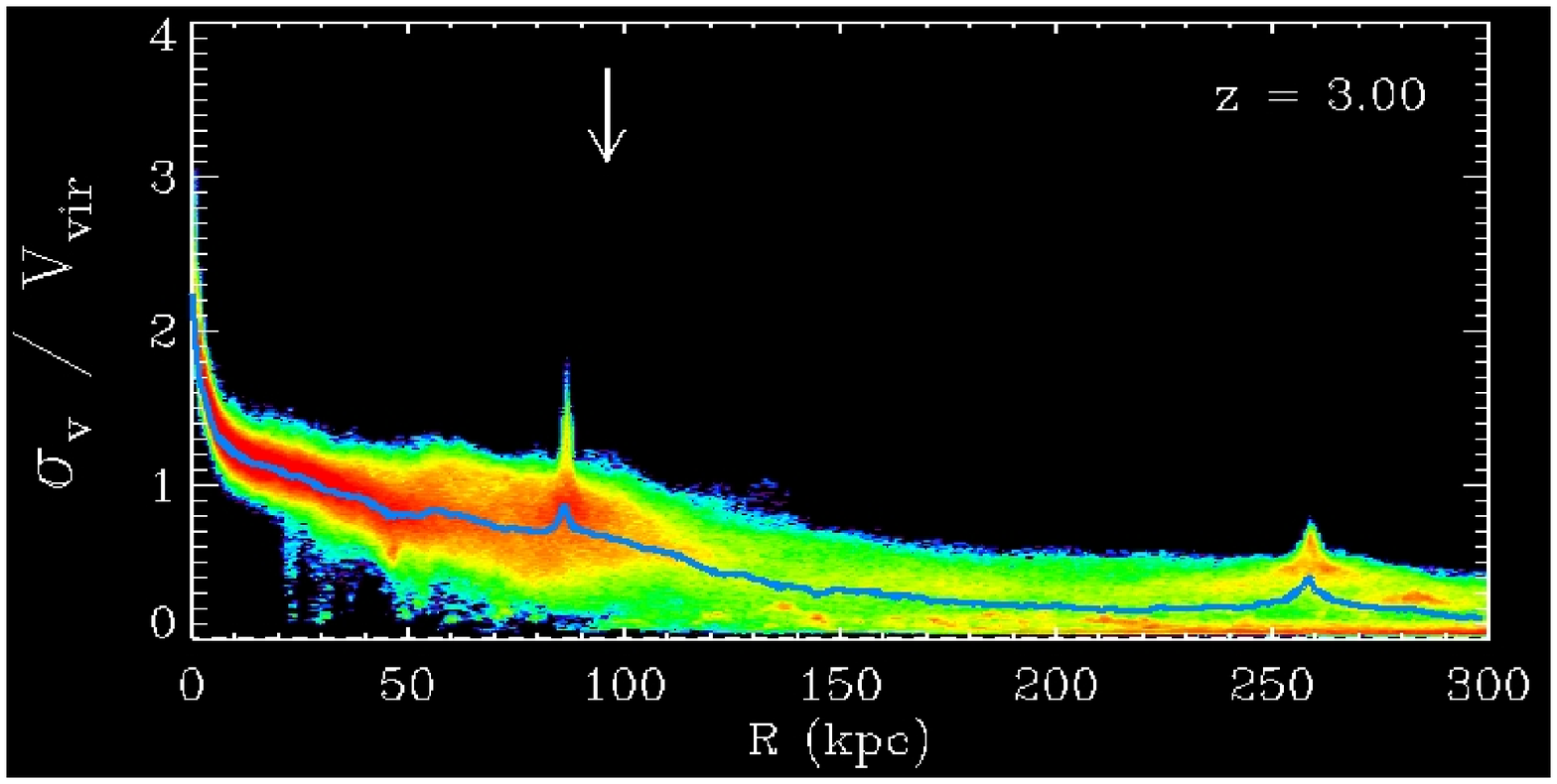}
\includegraphics[angle=0,scale=0.408]{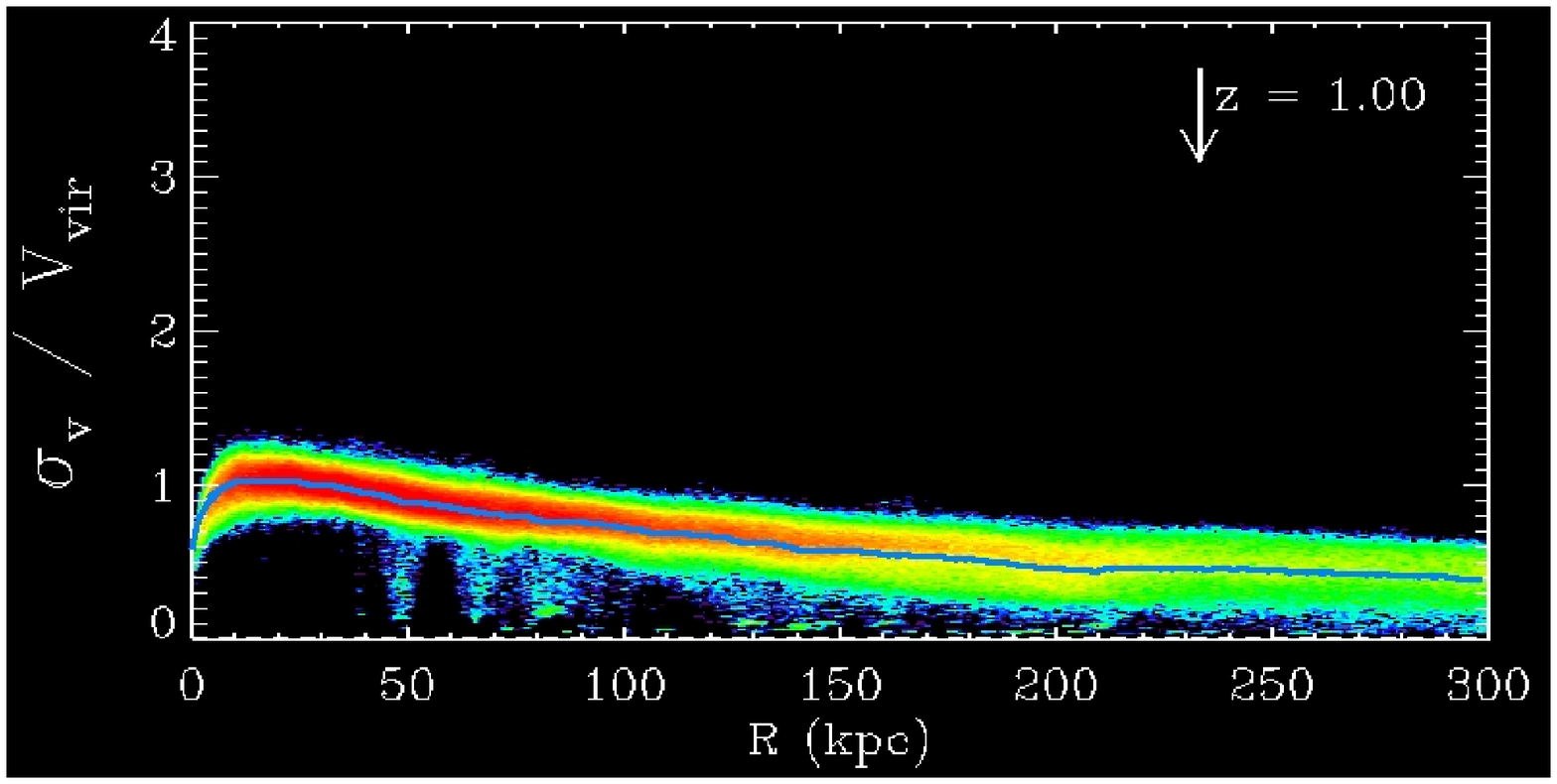}
\includegraphics[angle=0,scale=0.408]{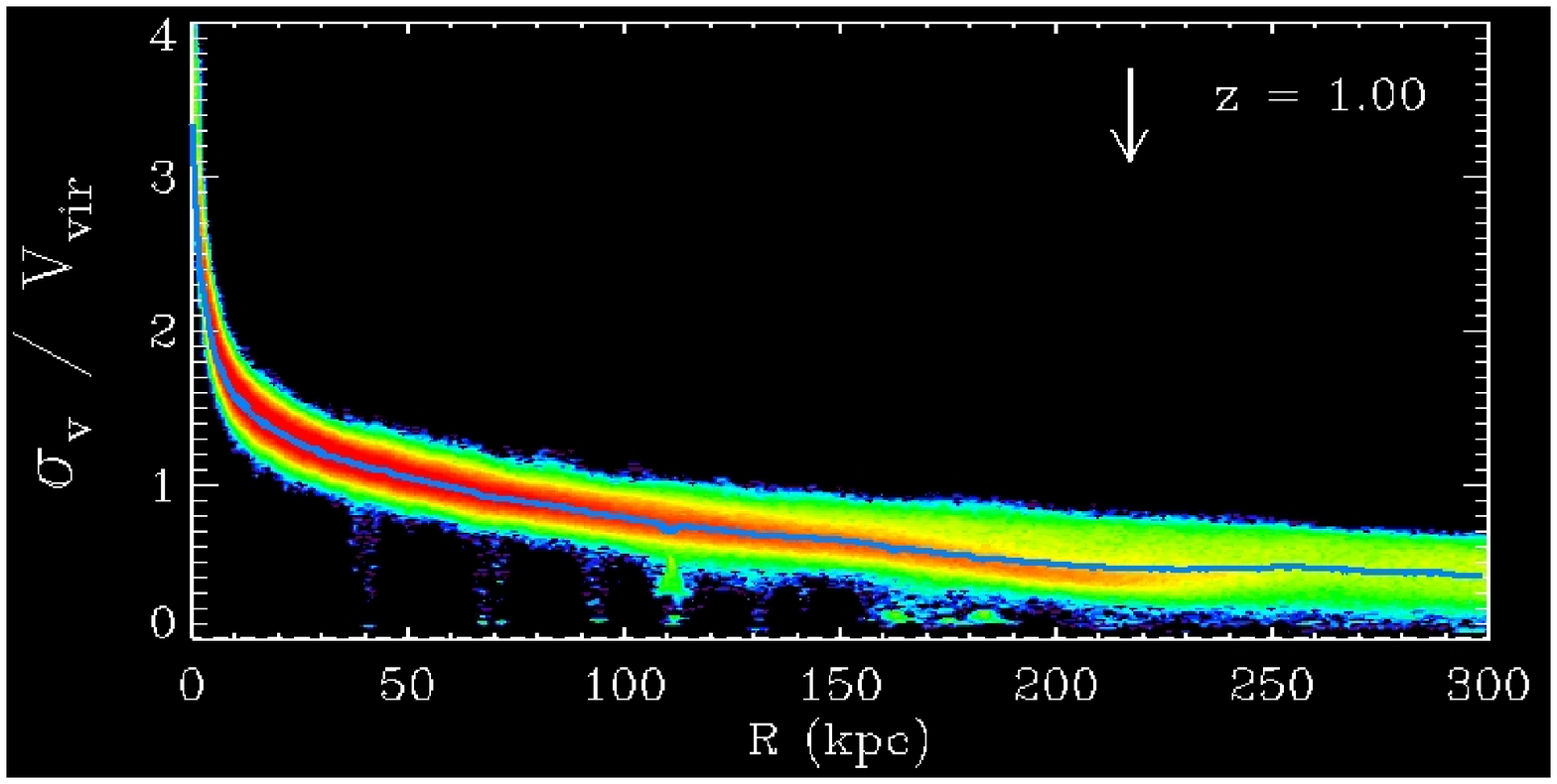}
\includegraphics[angle=0,scale=0.408]{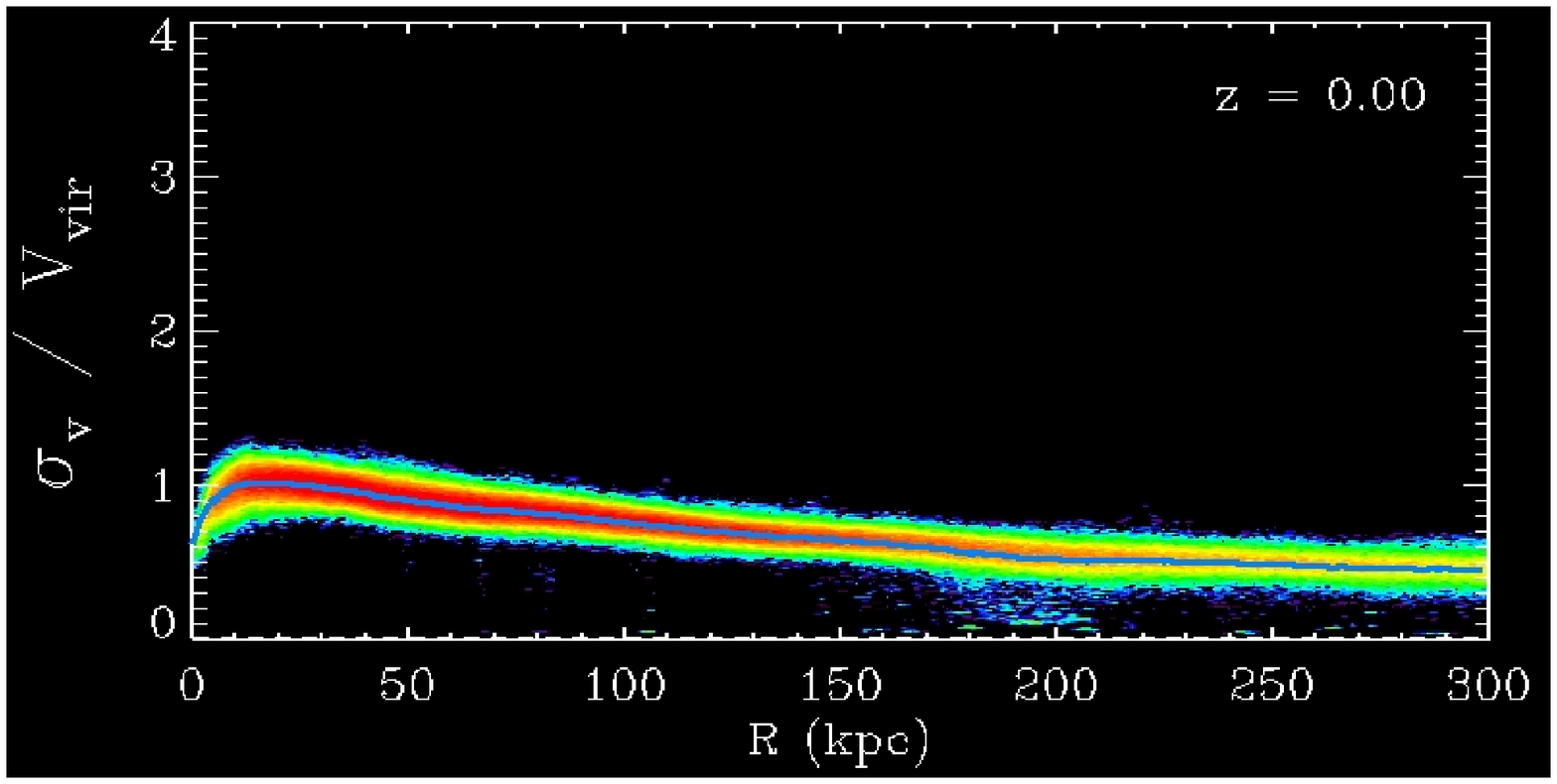}
\includegraphics[angle=0,scale=0.408]{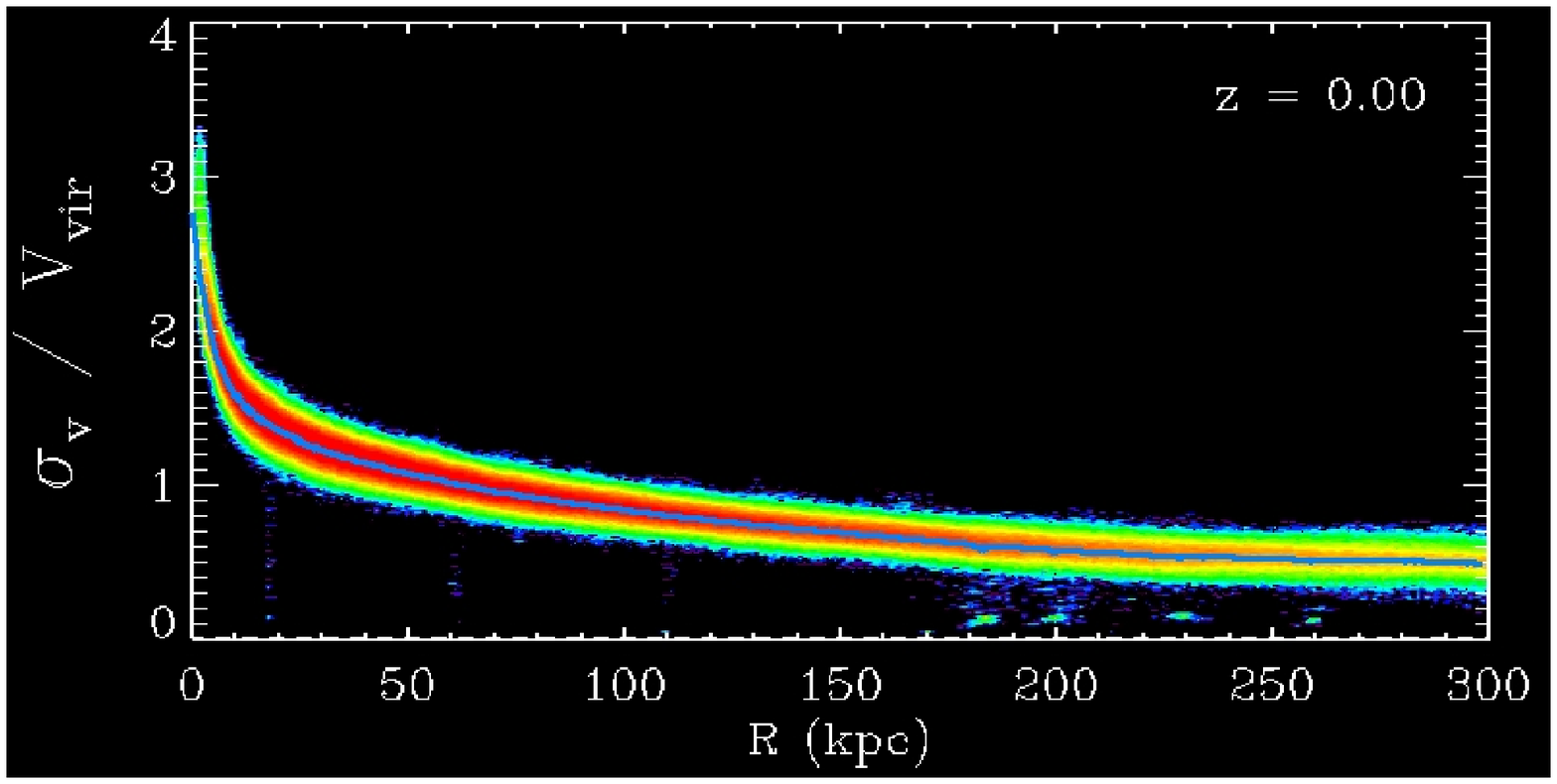}
\end{center}
\caption{DM dispersion velocity in PDM (left) and BDM (right) models
  at various $z$. The colors correspond to DM particle densities.  The
  y-axis is normalized by the circular velocity at \rvir{} at $z=0$.
  The blue lines show the average values of dispersion velocities at
  each $R$ and the colored width represents a $1\sigma$ dispersion
  around the mean. The velocity axis is normalized by $v_{\rm vir}$
  --- the circular velocity at \rvir.  }
\end{figure*}

Snapshots of tangential velocities of DM particles are shown in
Fig.~12.  During the initial expansion these velocities `cool'
down. This distribution is symmetric with respect to $v_\phi=0$ up to
$z\sim 9-10$, when the growth of the inhomogeneities feeds $v_\phi$
with positive or negative asymmetries at small $r$, during the major
mergers epoch.  While each major merger forces positive or negative
asymmetry at the center, the velocities relax to a symmetric
distribution thereafter. Not only major, but also minor mergers can be
easily traced in this diagram.  By $z\sim 2.5$, this diagram achieves
a higher degree of symmetry which is only perturbed by merger
events. The $v_\phi$ distribution widens with $r$ when entering the
virialized halo radius.  The high degree of symmetry between the
number of prograde and retrograde circulating particles is very
important in order to understand the dynamical state of the DM halo as
well as its ability to interact with the baryonic matter at its
center.

\begin{figure*}[ht!!!!!!!!!!!!!!!!!]
\begin{center}
\includegraphics[angle=0,scale=0.9]{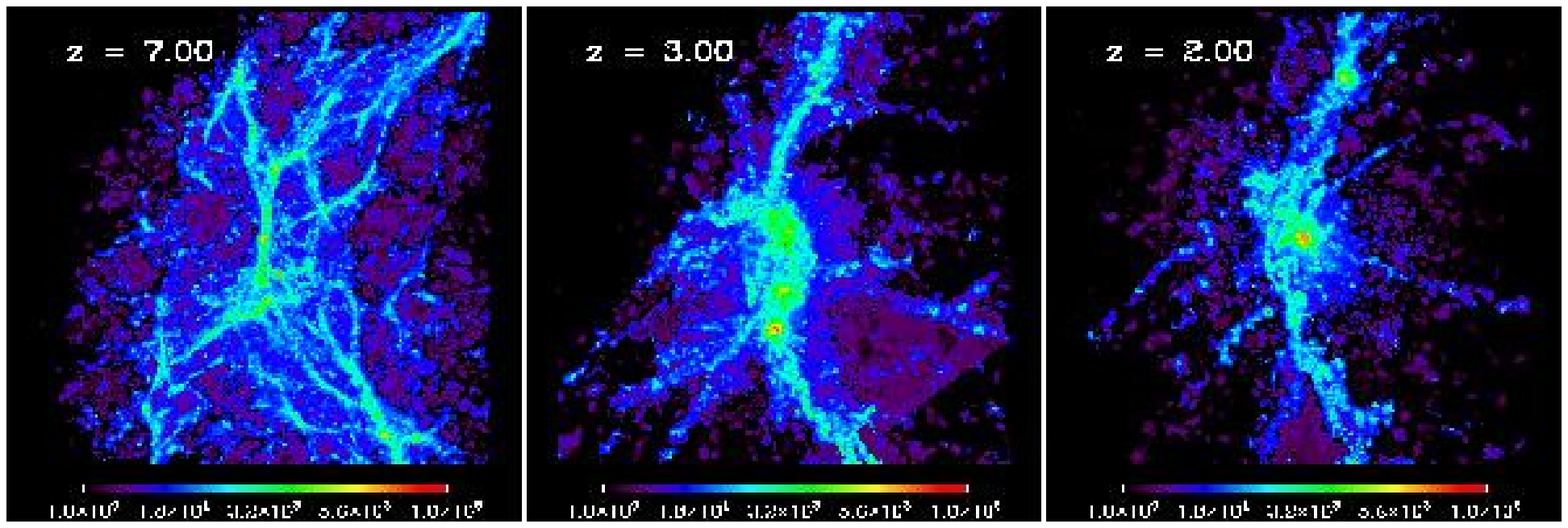}
\end{center}
\caption{Evolution of gas temperature within the central $\sim
  400$~kpc (as of $z=0$) of the prime halo at $z=7$, 3 and 3.
}
\end{figure*}
\begin{figure}[ht!!!!!!!!!!!!!!!!!!]
\begin{center}
\includegraphics[angle=0,scale=0.43]{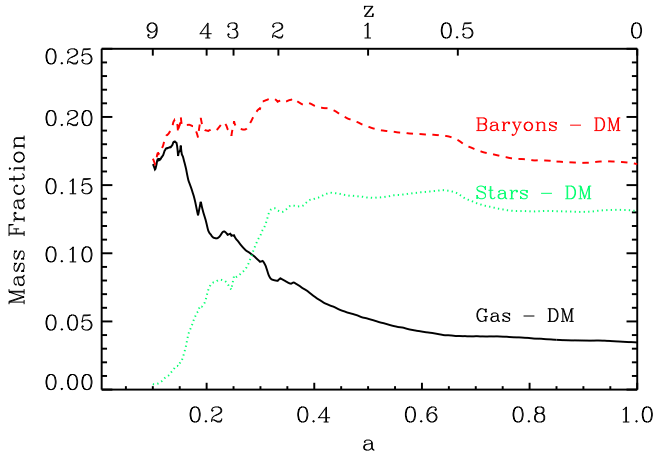}
\end{center}
\caption{Evolution of gas, stars and baryonic fraction within \rvir\
of the prime halo. The ratios are given with respect to DM.  }
\end{figure}
\begin{figure}[ht!!!!!!!!!!!!!!!!!!]
\begin{center}
\includegraphics[angle=0,scale=0.41]{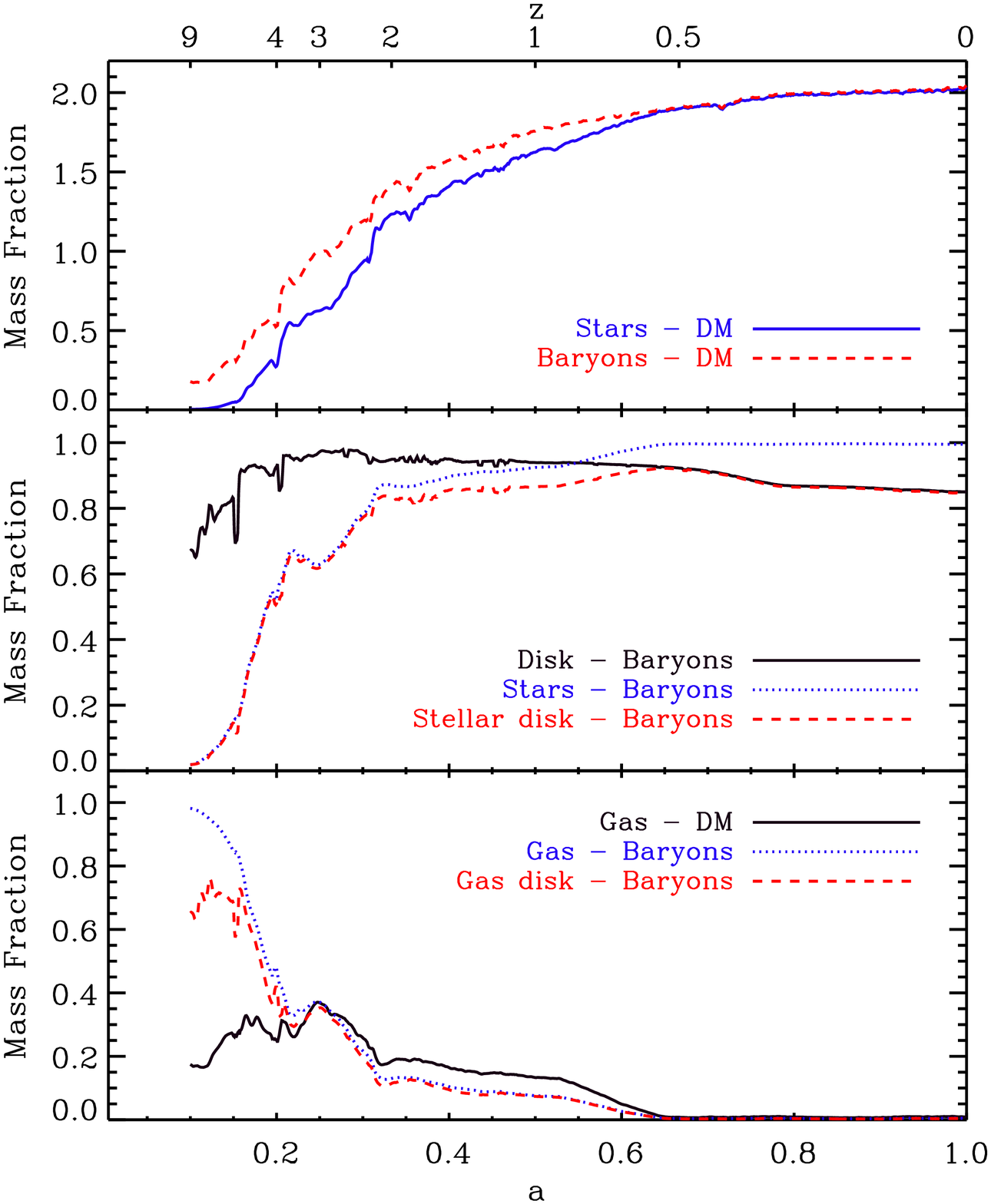}
\end{center}
\caption{Evolution of mass ratios within the central 10~kpc: stars/DM
 and baryons/DM (top), disk/baryons, stars/baryons and stellar disk/baryons
  (middle), and gas/DM, gas/baryons and gas disk/baryons (bottom).  }
\end{figure}

The white lines in Fig.~12 correspond to the circular velocities in
the main halo as a function of $r$ and are calculated for a 
spherically-symmetrised halo at $z$.
The halo is substantilly triaxial at all radii and the shown circular
velocities provide a bad approximation to the local velocity field in
that
they underestimate velocities within \rvir{} and overestimate them
outside.  The overall symmetry in the $r-v_\phi$ diagram confirms that
there is very little net circulation of the DM within the halo. We
also noted that the halo figure does not tumble and is oriented along
the main filament which feeds its growth over the Hubble time.

The first time entering subhalos inside \rvir{} move close to the
$v_\phi=0$ line, as they follow the cold stream from one of the
filaments and have a very low angular momentum with respect to the
prime halo. Only at small $r$ their $v_\phi$ become appreciable.

Residual maps in the $r-v_\phi$ plane constructed from Fig.~12
snapshots along the lines outlined above provide very similar results
on the estimated mass fraction in the substructure (Fig.~11), and we
avoid from displaying it here.

\subsection{Halo Evolution: Dispersion Velocities}

The evolution of DM in the presence of baryons differs in the dispersion
velocity map. These differences become visible with the formation of
first bound condensations (Fig.~13). The subhalos are much more bound
and their internal dispersion velocities reflect this clearly (e.g.,
spikes). The low velocity dispersions, $\sim 10~{\rm km~s^{-1}}$,
describe the conditions inside the filaments penetrating the
halos. The subhalos in the filaments also exhibit low internal velocities. 
As the inflow penetrates to smaller $R$, the smaller clumps
are obliterated. At higher $z$, the clumpy nature of the inflow is still
visible, but at low $z$ becomes much less obvious.

Next, dispersion velocities display a divergent behavior in the
center, $R\ltorder 20$~kpc. While they peak in the BDM, they decline
in the PDM (see also Romano-Diaz et al. 2008b). The turnover radius 
corresponds to \rts{} in the BDM model.

\section{Results: Kinematics of Baryon Buildup}

Baryon influx within \rvir{} proceeds along the filaments and is clumpy
(Paper~II).  The gas temperature gradients become visible at high $z$
due to shocks as the gas enters the filaments, and due to the 
virialization within subhalos, delineating the clumpiness along the
filaments. By $z\sim 12$, $T$ reaches $\sim 10^4$~K in the most
prominent virialized centers. The web structure becomes `illuminated'
by $z\sim 8$. Fig.~14 shows the gas $T$ within the central 400~kpc of
the prime halo. At later $z$, the feeding filaments are clearly
visible because of their lower gas temperature --- the smooth
(i.e., unresolved) gas
component accompanies the subhalos as they penetrate to the center
along the filaments like beats on the wire. The filaments extend from
the computational boundaries to the very central region. The typical
inflow velocity along the filaments scales roughly as the free-fall 
velocity, i.e., $v_{\rm ff}\sim R^{-1/2}$ within \rvir{} and is measures 
in 100s~${\rm km~s^{-1}}$ in the central regions. The filaments weaken 
after $z\sim 2$.

\begin{figure}[ht!!!!!!!!!!!!!!!!!!]
\begin{center}
\includegraphics[angle=0,scale=0.45]{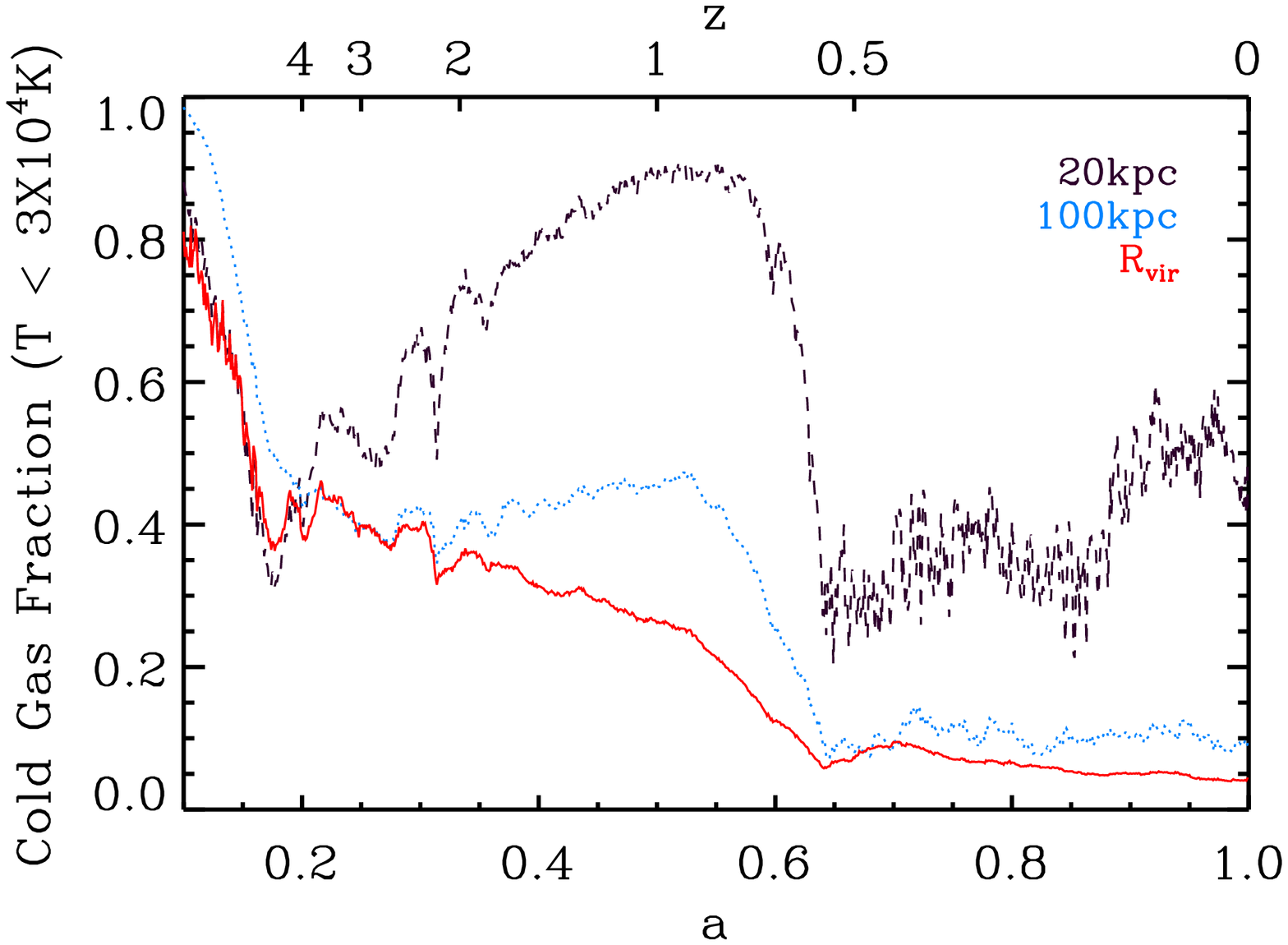}
\end{center}
\caption{Evolution of cold, $T\ltorder 3\times 10^4$~K, gas mass fractions
  (of the total gas within $R$) within the central 20~kpc,
  100~kpc and \rvir{} of the prime halo. The fractions are normalized
  by the total gas mass within the same radius. Note, that the 100~kpc
  curve lies outside \rvir{} for $z\gtorder 3$.  }
\end{figure}
\begin{figure}[ht!!!!!!!!!!!!!!!!!!]
\begin{center}
\includegraphics[angle=0,scale=0.45]{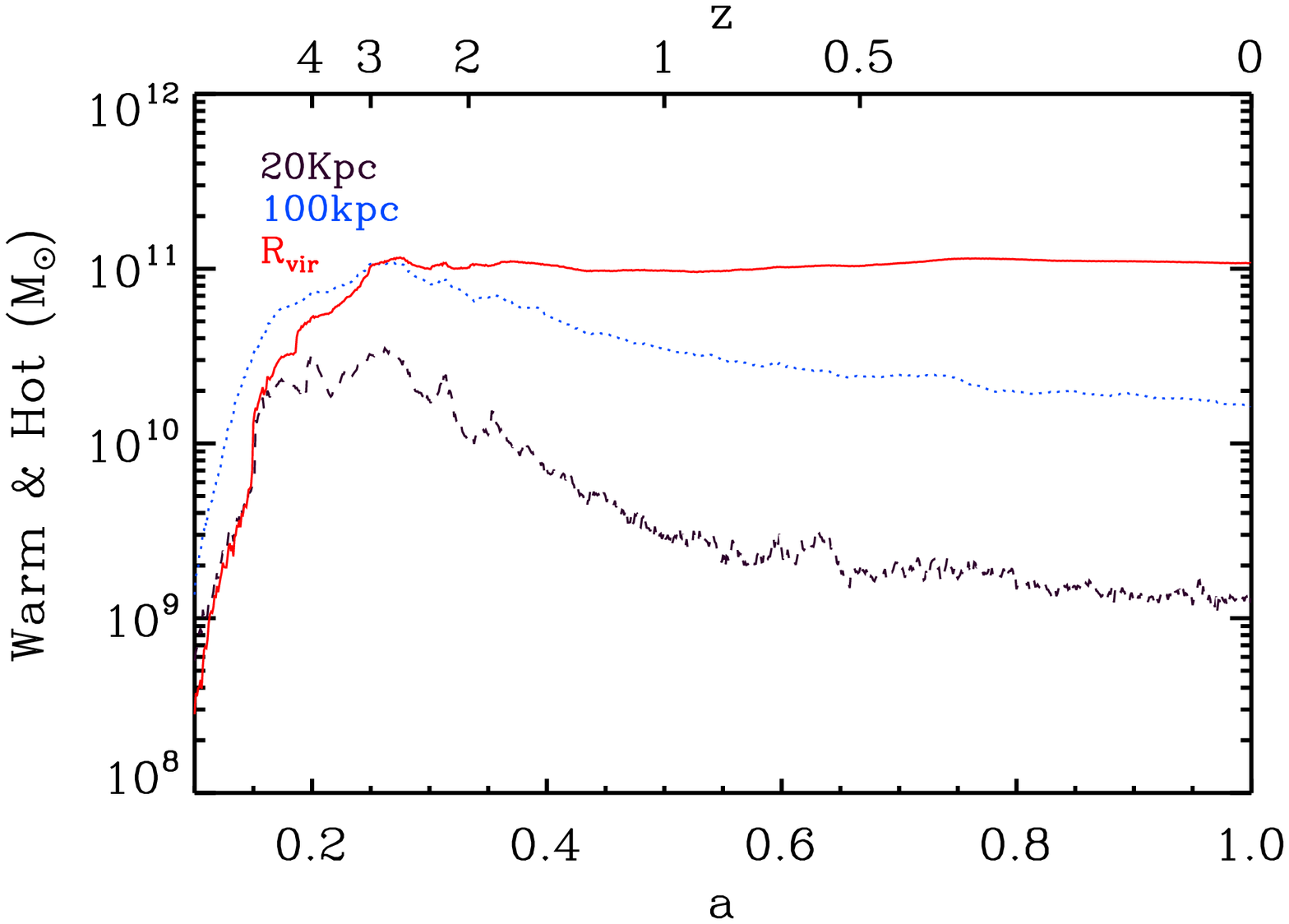}
\end{center}
\caption{Evolution of warm and hot, $T\gtorder 3\times 10^4$~K, gas
mass within the central 20~kpc, 100~kpc and \rvir{} of the prime halo. 
Note, that the 100~kpc curve lies outside \rvir{} for $z\gtorder 3$. }
\end{figure}
 
Next, we calculate the baryon fraction with respect to the DM, 
$f_{\rm bary}$. 
Initially, in the WMAP3 Universe $f_{\rm bary}\sim 17\%$. 
The distribution of mass among the components within \rvir{} as a
function time is shown in Fig.~15. At $z=0$, the DM mass in the BDM 
prime halo is \mvir$\sim 3.2\times 10^{12}~\msun$,
the gas $\sim 1.1\times 10^{11}~\msun$ and stars 
$\sim 4.2\times 10^{11}~\msun$. Hence the final $f_{\rm bary}\sim 17\%$
is close to the initial ratio. Moreover, most of the time 
$f_{\rm bary}$ is even higher, reaching its peak of $\sim 22\%$ toward the
end of the major merger epoch. We discuss the implications in \S6.
Here we note that the spheroidal {\it stellar} halo component
which forms as a result of subhalo dissolution in the central region
has a steep density profile of $\sim (R/10~{\rm kpc})^{-3.5}$.
Its mass contribution to the halo is not large.

The evolution of the mass ratios within the central 10~kpc is shown in
Fig.~16. This is the region which hosts a stellar/gaseous 
disk\footnote{We define the gas (stellar) disk by determining the rotation
axis of gas (stars) within the central 5~kpc. Within the equatorial
plane we determine the surface density, and radial and vertical density
distributions, etc.} whose
detailed properties are outside the scope of this work (e.g., 
Romano-Diaz et al. 2008c). The stellar/DM ratio reaches unity by 
$z\sim 2$ and is just below 2 at $z=0$. The baryons are concentrated 
in the disk by $z\sim 3$, but disk/baryon ratio declines to about 0.8 
at later times, reflecting the formation of a spheroidal baryonic 
component of stars and some hot gas. By $z\sim
0.5$, the central 10~kpc loses most of its gas component (except in the
central 1--2~kpc), which is associated with a decline of the SF in the 
disk to $\sim 1\msun$. The disk itself appears nearly gasless by the 
end.

We note that the {\it initial} disk morphology reflects the prevailing
asymmetry of the background DM gravitational potential which is
nonaxisymmetric. This leads directly to the formation of an asymmetric
gas disk which gives rise to the stellar/gaseous bar during the first
Gyr (Romano-Diaz et al. 2008c).  The disk is rebuilt during the major
merger epoch and experiences a number of interactions with the
subhalos. A long-lived stellar bar is triggered by one of the prograde
encounters. This bar drives a radial gas inflow in the disk,
fueling the SF activity in the central kpc or so.

Fig.~17 shows the evolution of cold, $T\ltorder 3\times 10^4$~K, gas
fraction (of the overall gas content) within the inner 
20~kpc disk region, 100~kpc and \rvir. The cold gas inside 
\rvir{} is
decreasing monotonically until $z\sim 0.5$, then levels off for the
rest of the simulation, reflecting the gas accretion rates along the
filaments. At smaller characteristic radii, this gas fraction behaves
differently. Within 100~kpc it is steady till $z\sim 0.8$, well
beyond the major merger epoch, then sharply declines and
stabilizes. Most interestingly, the cold gas fraction within the disk
region grows dramatically from $\sim 30\%$ at $z\sim 5$ to about
$90\%$ at $z\sim 0.8$. A sharp decline to $\sim 25\%$ is well
correlated with the cutoff in gas accretion via filaments and a splash
in the inflow rate of subhalos (Romano-Diaz et al. 2008b). Toward
$z=0$, this fraction is around $45\%$, but the overall gas content in
this region is small, already driven out by the minor interactions
with the substructure. Hence about 2/3 of the cold gas residing in the
disk at $z\sim 0.7$ is driven out by the interactions. The remaining
1/3 is largely used up in star formation.

The behavior of the hotter, $T\gtorder 3\times 10^4$~K, gas which is
not confined to the disk, but constitutes the spheroidal component
mimics the evolution of the DM there in many ways (Fig.~18). While the
total amount of this gas inside \rvir{} (which grows with time) is 
nearly constant in time, it is
steadily declining after $z\sim 3$ at any fixed radius. This decline 
extends beyond the major merger epoch. 
If one relies on Figs.~16 and 17 to estimate the heating effect of
mergers, the integral contribution of minor mergers in heating this 
gas and driving it out appears similar to that of the major mergers.
   
\begin{figure*}[ht!!!!!!!!!!!!!!!!!!!!!!]
\begin{center}
\includegraphics[angle=0,scale=0.35]{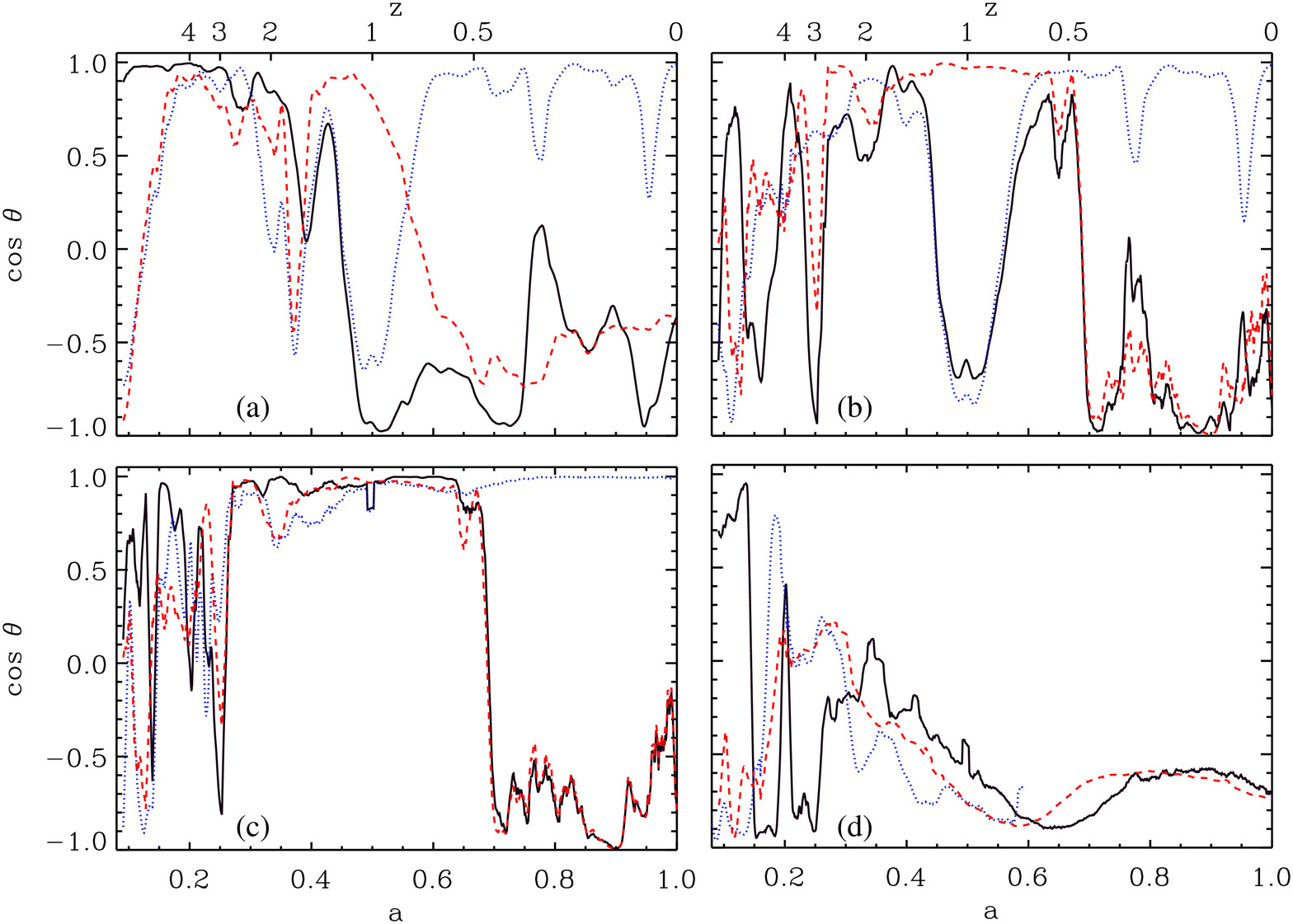}
\end{center}
\caption{Correlations between angular momenta of 
the DM, gas and stars within (a) \rvir, (b) \rvir{} (DM) and 8~kpc 
(gas and stars), and (c) 8~kpc (DM, gas and stars). The angle $\theta$
is defined between the corresponding $J$'s. The colors are those of
DM--gas (black, solid), DM--stars (blue, dotted) and stars-gas (red, dashed).
(d) The orientation of $J$'s within the central 8~kpc with respect to 
the inertial frame. The colors are those of DM (black, solid), gas 
(blue, solid) and stars (red, dashed). 
 }
\end{figure*}

We have estimated the stellar mass distribution in the final
disk, using vertical half-thickness of 2.5~kpc, and find that the disk 
contribution to the rotation curve dominates within 12~kpc. The 
final disk-to-virial mass ratio in the halo is $\sim 7\%$. Detailed
study of disk dynamics is outside the scope of this paper, however we
make a few related comments. First, the disk has lost most of its gas,
$\sim 2/3$,
during the late quiescent epoch when it was experiencing frequent 
interactions with the subhalos and the gas accretion has ceased. (The 
rest of the cold gas was converted into stars.) This loss of the cold
gas is reflected by the decline in the SF rate by a factor of 10 (e.g.,
Fig.~5 of Romano-Diaz et al. 2008c).
The {\it decline in SF rate by a factor of $\sim 10$ is a consequence,
of this interaction process which drives the gas away from the disk}. 
Hence, the dramatic fall in the SF rate does not result from the gas 
conversion into stars. Second, after $z\sim 0.5$ the disk has
been also losing
its stellar component which was heated up due to the same interactions.
This can be see in Fig.~16 (mid-frame), where stellar (and total) 
disk/baryon ratio within the central 10~kpc declined. We have not 
attempted the disk-bulge-bar decomposition, but use the mass ratio of 
the {\it stellar} spheroidal component to stellar disk within the disk 
radius. This ratio has increased with time from 0.1 at $z\sim 2$, to 
0.2 at $z\sim 1$, to 0.7 at $z=0.5$. In tandem with the gas loss from 
the disk, this hints at morphological evolution from late Hubble type to 
intermediate and early type disk.
 
A new and potentially interesting mechanism of truncating the stellar 
disk was noticed. At various times, highly inclined and massive gaseous
rings have been spotted to form around the disk, with radii 
$\sim 15-20$~kpc. They did not evolve into polar rings 
(e.g., Sparke 2002), appear short-lived and related to the gas 
accretion through the filaments. No attempt was made to quantify
their effect on the disk dynamics, but they are expected to exert 
substantial perturbarions on the {\it direction} of the angular momentum
vector in the disk region (see below).

We have calculated the evolution of angular momentum in the gas and
stars and compared it with that of the DM at various spatial scales.
Prior to $z\sim 3$, the orientation of the total $J_{\rm DM}$ 
correlates nicely with that of the total $J_{\rm gas}$ inside \rvir,
while this radius is well below 100~kpc (Fig.~19a). After $z\sim 1$, 
the gas 
appears to counter-rotate at high inclination to the DM globally. 
At this time almost all the gas is hot, $T\gtorder 3\times 10^4$~K, 
within \rvir. The global gas and stellar $J$ are aligned only between
the major mergers and around $z\sim 1-1.7$. This alignment occurs when
the gas disk is dominated by the cold gas component. 

When global (i.e., within \rvir) $J_{\rm DM}$ is contrasted with
$J_{\rm gas}$ and $J_{\rm gas}$ within the disk region of inner 
8~kpc (Fig.~19b), the correlation between DM and stars, and DM and 
gas becomes 
much better after the major mergers and before the cold gas is
ablated from the disk. When all $J$'s are limited to
within the central 8~kpc (Fig.~19c), the correlations between the 
components
become even better. Interestingly, the cold gas that is
concentrated within the central 1--2~kpc after $z\sim 0.5$ 
anticorrelates with DM and therefore anticorrelates (i.e.,
counter-rotates) with stars as well.

After the stellar disk formation,
its rotation axis experiences an oscillatory motion within 
respect to the fixed inertial system of coordinates and is closely 
followed by the DM axis within the same region. This trend is 
interrupted by the mergers but always returns (Fig.~19d).
  
\section{Discussion and Conclusions}

We have compared the buildup of DM halos with and without baryons, BDM
and PDM respectively, from identical cosmological initial
conditions. As a template, we choose to follow the evolution of massive,
Milky Way-type halos which emerge from their major merger phase early,
by $z\sim 1.5$, and remain isolated thereafter. We have followed the 
evolution of characteristic quantities in the halos, i.e., \rts{} 
($\equiv \gamma$\rvmax$\sim 0.46$\rvmax), \rvmax, \rvir, the
radial density distribution, the bound fractions, the phase-space
distribution function, figure tumbling, and the full and specific
angular momenta. Furthermore, we have
analyzed the halo relaxation in the configuration and phase spaces and
again compared how this process is affected by baryons. We find that
following the evolution of DM accretion in the $R-v_{\rm R}$ and $r-v_\phi$
planes allows one to observe and quantify the development of 
substructure within the halo --- the subhalos, tidal tails and
streamers. Finally, we comment on the baryon buildup within the halo
(see also Romano-Diaz et al. 2008b,c).  The quantitative evolution of
the substructure is further analyzed in Paper~II.

The resulting PDM and BDM halos have been compared
to the halo distribution in the 
\mvir$-z_{\rm form}$ plane for the Types~III and IV halos of McBride et al. 
(2009), where $z_{\rm form}$ is the halo formation time using
McBride et al. definition.
They deviate in about $2\sigma$ from the mean that was obtained
by compiling an extensive catalog of $\sim 500,000$ halos from the 
Millenium simulation. Types~III and IV, together are the most significant
population of halos. Moreover, our halos lie within $1\sigma-2\sigma$ 
from the mean in Li et al. (2007) who studied the Mass Accretion Histories
(MAHs) of DM halos in large cosmological boxes. This lowers our halos 
probability accordingly, but they remain plausible. 
In the mass range
of $10^{12-13}~{\rm M_\odot}$ in Li et al., the probability distributions of halo
properties sample objects with only a few 100 particles, 
making it
difficult to reliably estimate the width of the distribution. But it
seems that our DM halos grew faster than MAHs from Li et al. show, by a 
factor of $\sim 2$, during the major merger epoch. The reason for this
is explained in \S3.1 and has no effect on the actual evolution.
 
As a next step, we attempt to understand our results presented
in \S3--5.  The prime halo accumulation in these simulations happens
in the loose field rather than in the dense cluster environment because
the forming halo is embedded in the region whose overdensity is
zero. The number of major mergers in the system is determined by the
random component of the CRs.  We first compare the key parameters of
PDM and BDM model halos. The halos have nearly the same mass within their
virial radii, but the large-scale mass distribution in both models
differs in its degree of triaxiality. As expected, the BDM model ends
up much less triaxial, being nearly axisymmetric and somewhat
flattened. The evolution of the innermost mass distribution diverges
in both cases as well. While the PDM halo quickly acquires the NFW
cusp with $\rho\sim R^{-1}$ and a stable \rs$\sim 28$~kpc after the 
epoch of major mergers,
the BDM halo develops an isothermal cusp with $\rho\sim R^{-2}$, 
bypassing the NFW cusp
(Romano-Diaz et al. 2008b). The isothermal cusp is stable between
$z\sim 1-4$ and is gradually erased thereafter, leading to the
formation of a flat DM core in the central 2--3~kpc. The reason for
this behavior is two-fold: first, the adiabatic contraction is
initiated by the dissipative baryons, dragging the DM inwards. The
major mergers do not affect the log$\rho-$log$R$ slope of --2 for
the DM mass distribution. However, the subsequent minor mergers 
penetrating the
inner 20--30~kpc are accreted in groups of a few because the subhalos
cluster already during their motion along the filaments.
Not only their dynamical friction affects the isothermal DM cusp
but they also have a profound effect on the baryons in the cusp, when
the cold disk gas is ablated and, together with the inner halo hot
gas, is driven out.

The diverging evolution of PDM and BDM halos is also demonstrated by
their concentration parameter $\tilde{c}$ (Fig.~4). The latter halo is much
more concentrated because of the smaller \rts$\sim 15$~kpc vs \rts$\sim 
28$~kpc in the PDM. The abrupt change in the
growth of $\tilde{c}$ after $z\sim 0.5$ is closely associated with the minor
mergers and clustering of accreted subhalos, coming from the filaments.

In phase space, we find that the phase space density, $Q(R)$, of
the DM in the BDM model cannot be fit by a power law, unlike in the 
PDM model. The differences appear in the central $\sim 30$~kpc, as
a downward trend
after $z\sim 4$. After $z\sim 0.5$, the central $Q$ drops even faster
and $Q(R)$ flattens within the central few kpc (Fig.~5).
This can be related to the formation of an isothermal density cusp 
there between $z\sim 4-1$ and its gradual flattening from inside out 
after this time. This means that the entropy per DM
particle, defined as $\sigma^2_{\rm DM} \rho^{-2/3}$ (e.g., Hoffman et
al. 2007), increases in the central region of the BDM halo compared
to the PDM.
 
Both PDM and BDM halos show very little figure tumbling with respect
to any axis (Fig.~6), based on $\sim 900$ snapshots. This happens 
despite some DM particles acquiring
substantial angular velocities with respect to the halo CoM. The
cosmological $\lambda$ falls within the limits normally attributed to
the halo angular momentum, $\lambda\sim 0.01-0.1$. 
The angular momentum of the halo during its assembly is channeled
into the internal circulation of DM particles and not into the
tumbling of their orbits in any collective fashion. By internal
circulation we mean $J$ associated with individual orbits which may or
may not correlate among themselves. If they do, the halo will possess
a net $J$ (i.e., $\lambda$), but its figure can remain stationary 
nevertheless.

The halo figure tumbling was claimed to be insignificant, if evolved
in a small isolated box or for cluster mass halos (e.g., Dubinski
1992; Bureau et al. 1999).  Heller et al. (2007) found that, for a
large number of galactic halo models with and without baryons evolved in
isolation from high $z$, the resulting halo figures tumble exceedingly
slow, with an average $\Omega_{\rm h}\sim 0.2~\kmskpc$ around their
minor axes. This agrees nicely with Bailin \& Steinmetz (2004)
estimates from 4 snapshots for halos extracted from cosmological
simulations. A departure from axial symmetry, especially during
the early epoch, taken in tandem with the exceedingly slow tumbling of
the halo figure has important {\it dynamical} consequences for the growing
baryonic disk.  The simplest explanation to the absence of tumbling in
the PDM and BDM halos comes from the tidal effects of the large-scale
DM filaments. This is confirmed by the stable orientation of the halo
major axis with respect to the filament. The same effect is more
difficult to understand in the context of isolated halos with angular
momentum inserted via initial conditions (Heller et al. 2007).

The evolution of angular momentum associated with spherical shells
within DM halo is complex, but the observed increase in $J($\rts$)$ and
$j($\rts$)$ of the BDM halo compared with the PDM is a signature of an 
angular momentum transfer from the massive disk in the innermost halo.  
We also detect changes in the position angle of the total angular 
momentum ${\vec {\bf J}}($\rvir$)$ of the halo, both in the PDM and BDM 
models, while their value is stable. The change in the orientation of 
$J$ is much easier to achieve than the change in its value. 

Potentially interesting are the correlations between the angular momenta
of baryons (stars and gas) and DM, shown in Fig.~19 as an angle between
the pairs of $J$'s. On the scale of \rvir, $J_{\rm DM}$ 
is aligned with $J_{\rm gas}$ untill $z\sim 3$, and largely 
anticorrelates (anti-aligned) at $z\sim 1-0.5$ (Fig.~19a). The
picture differs when the disk region baryons are correlated with the 
DM within \rvir{} --- abrupt changes occur during the mergers (Fig.~19b).
The best correlation is obvious when all components are compared 
within the disk region --- the DM and gas $J$'s are oriented
similarly here between $z\sim 3-0.5$ and then abruptly anticorrelate
(counter-rotate) at later times, when the cold gas is limited to the 
central kpc (Fig.~19c). 

Of course no correlation is maintained between $J_{\rm DM}$ within 
\rvir\ and $J_{\rm stars}$
within the stellar disk during mergers. Even during
the quiescent epoch, $z\sim 1$, the disk experiences a flip-flop
(i.e., from corotation with the halo DM to counter-rotation and back, 
Fig.~19b). This can be most probably related to the interactions with 
the streamers which survive for a long time, see below.

An important issue when analyzing the growth of a DM halo is how the
DM and baryons are deposited within its volume --- this brings us to
the question of internal relaxation. The halo experiences
virialization --- a particular relation between its kinetic
and potential energies is achieved. This process is typically
associated with a time-dependent gravitational potential. Clearly,
the halo acquires a degree of virialization only asymptotically,
as the rate of major mergers subsides. While virialization
proceeds on a dynamical timescale, its incompleteness is related to
the halo being an open system, accreting from its environment.

The relaxation process extends beyond virialization and continues both
in the configuration and phase spaces. In the former, we ask how was
the halo assembled.  Did its central part, say within \rts, 
form early
in its evolution, e.g., during the major merger epoch, as advocated by
Wechsler et al. (2002), and does the presence of baryons make any
difference in this process?  How quickly the substructure evolves
within \rvir? (By substructure we mean subhalos, their tidal tails and
streamers, as defined in \S4.) In the latter, we ask how long the
position-velocity correlations persist after dissolution of the bound
structures (subhalos) within the halos, and again, do baryons have any
effect on this process?

We start by comparing some details of radial mixing of DM
halos and a non-singular `isothermal' sphere (NSIS). The fractions of 
`bound'
particles --- those limited in their radial motion to within a
specific radius by their energy are shown in Fig.~7, where $R$-axis
was normalized by \rvmax. After the epoch of major mergers, the
fractions of bound particles, $\eta$, within $R$ stay constant with time. We
see that the fraction of bound particles within $\sim $\rvmax{} is
substantially higher in the PDM and BDM halos compared to NSIS, while
outside this radius it is nearly the same.  Hence, particles of the
NSIS found within \rvmax, will perform larger radial excursions than
those found in the modeled halos. In this region, the BDM and PDM
halos are more bound than NSIS. But even with this higher 
fraction in the BDM halo $\eta$  is still only $\sim 23\%$ within \rvmax,
$\sim 15\%$ within \rts, declining sharply (faster than in the
NSIS) within \rts, and even faster within the NFW cusp of the PDM and
the flat density core of BDM. Inside the cusp, 
$\eta$ drops well below $1\%$. This means
that the majority of particles in the PDM and BDM halos, at virtually
every radius inside \rvmax, are not confined within their current 
$R$, especially
in the central regions. Hence we cannot confirm that particles
accreted during the major mergers form the core of DM halos. The part
of the PDM and BDM halos inside \rvmax{} must be well mixed.  Energy
considerations provide a relatively weak constraint on the motions of DM
particles within this region.

The fraction of {\it bound} particles, albeit small, within \rvmax{} 
assemble within the major merger epoch, in PDM and BDM models. How can
this be reconciled with the DM density profiles which are believed to
form in the early stage of evolution? While the fractional
contribution of (minor) subhalos is insignificant within \rvmax{} in
both PDM and BDM, their fate differs substantially. The analysis of
Paper~II (see also Romano-Diaz et al. 2008c) shows that many of the
PDM subhalos are tidally destroyed before they enter \rts, while the
BDM ones, `glued' by baryons, survive the radial plunge. It is
important that the flow of the {\it unbound} DM contributes about
$80-90\%$ of the DM within \rts{} after the merger epoch, 
but {\it the net influx of this material is zero,} and this region 
appears to be in a steady state (Fig.~8). Most of the DM particles within 
\rts{} reside there for a short crossing time only.

Next, we analyze the relaxation of DM particles and their
substructures in the $R-v_{\rm R}$ and $r-v_\phi$ planes. The general
framework of halo assembly is that of the background filamentary
structure which triggers the DM and baryon flow.  As discussed in \S4,
this process is accompanied by violent relaxation whose efficiency
depends on the presence of substructure, and especially the residual
currents in the halo which continue to `mix' the halo environment even
in the absence of large-amplitude variability in the central
potential. Fig.~10 (and Animation Seq.)  phase diagrams
delineate the DM substructure in the various stages of its
evolution. The prime halo and subhalos are more centrally concentrated
with baryons, and the `fingers' are much more pronounced in this
case. Note the difference in the shape of $v_{\rm R}$ at small $R$ in
Fig.~10 and especially in the central 20~kpc between the PDM
and BDM models --- the DM potential is flat in the former and cuspy in
the latter model, as can be also be seen in the dispersion velocity maps
(Fig.~13).  The development of the flat core after $z\sim 1$ in the
prime halo does not reverse this trend. The absence of such cores in
the subhalos is most probably related to the numerical resolution.
Especially revealing are the later time frames which show that even
after the subhalos disintegration a substantial correlation in the
position-velocity diagram (i.e., streamers) remains within \rvir{} up
to $z=0$. Furthermore, as we discuss in Paper~II, shells of tidally
disrupted DM material travel outside \rvir{} and remain there at
present. After $z\sim 1$, we find that the outflowing material can
cross $\sim 1$~Mpc distance from the prime halo.

There are a few ways to quantify the substructure presence within
\rvir. Here we attempted to smooth the halo DM in $R-v_{\rm R}$ and
$r-v_\phi$ planes and use it as a (local) benchmark of the density
level in the halo. Subtracting it from the actual density map and
normalizing it by a reference mass \mvir, we obtained the excess mass
fraction associated with the substructure. This mass-conserving
procedure smears large density gradients in the phase-space. Fig.~11 
shows that the substructure becomes more
important with time in PDM and BDM models and is somewhat more
`visible' in the former model --- a result which seems to be
counter-intuitive. The simplest explanation of this phenomena lies in
the accumulation of streamers with time in both models and the more
efficient mixing in the BDM model.  The characteristic orbital time at
$\sim 200$~kpc is $\sim 2$~Gyr. Assuming that a few orbits are essential
for efficient mixing, the streamers that have been produced
before $z\sim 1$ are expected to disappear by $z=0$. Streamers
produced later on, most likely will survive to the present.

When analyzing the evolution of global and innermost mass fractions
(Figs.~15 and 16), we find that the baryon/DM fraction, $f_{\rm bary}$, 
within \rvir{}
increases during the major merger epoch and decreases during the
quiescent period. One should be cautious not to over-interpret
the final value of $f_{\rm bary}$ being so close to its initial
value. The crucial point here is the level of feedback from stellar 
evolution which drives the baryons out of the halo. Our models
have been fine-tuned and used the feedback values from Heller et al
(2007b). They are based on a large number of simulations where the
effects of the main free parameters in the SF have been investigated.
We have observed that an increase in the feedback results in smaller
bulge-to-disk mass ratios, while its decrease has led to extremely
bulge-dominated models.
The adopted values were those that led to the best correspondence of 
the final disks
to those observed. It is interesting that our choice resulted 
here in a near conservation of baryons within \rvir\ in the BDM model.

We now turn again to the baryon assembly within the DM halo. Most of the
baryon influx into the halo is channeled along the filaments (e.g., 
Keres et al. 2005; Ocvirk et al. 2008; Dekel et al. 2009) as a low
angular momentum gas (Fig.~14).  While we avoid here discussing the
disk dynamics (e.g., Romano-Diaz et al. 2008c), we do focus on the
parameters which characterize some aspects of the baryon accumulation
within the halo, especially within the central 10~kpc and a larger
disk region. Within 10~kpc, the
stars dominate over DM after $z\sim 2$ and the stars/DM mass ratio tends
asymptotically to $\sim 2$. The disk-to-total baryon mass ratio within 
this region tends to $\sim 0.8$ at present, which means that the 
stellar spheroidal contribution (i.e., bulge and innermost stellar halo) 
is $\sim 20\%$. 

After $z\sim 0.5$, the disk loses its gas
content, due to the highly non-steady influx of subhalos which ablate
the cold disk gas (see also Romano-Diaz et al.  2008b,c and 2009a for more
details). From the history of cold, $3\times 10^4$~K, and hot gas mass
fractions within the central 20~kpc (Figs.~17 and 18), we note that
the hot gas residing in the spheroidal component and corresponding
stars expand as a result of the subhalo heating of the immediate disk
region. Within the larger disk region, $\sim 15$~kpc --- 20~kpc, the 
ratio of stellar spheroidal
component to the total stellar disk mass increases with time,
from 0.1 at $z\sim 2$, 0.2 at $z\sim 1$, to 0.7 at $z\sim 0.5$. This
trend of increasing disk/spheroid stellar mass ratio is consistent
with evolution toward an early-type spiral galaxy. Similar evolution
was described by Naab et al. (2007) for one of their models, model E.

The `cold' baryons ending up in a disk in the BDM model
have resulted in $M_{\rm disk}/M_{\rm vir}\sim 7\%$. In comparison,
Xue et al. (2008) quote $M_{\rm disk}/M_{\rm vir}\sim 6.5\%$ for the Milky
Way galaxy, with $M_{\rm disk}\sim 6.5\times 10^{10}~{\rm M_\odot}$
and $M_{\rm vir}\sim 10^{12}~{\rm M_\odot}$, but higher estimates for
the MW disk exist as well, e.g., $8\times 10^{10}~{\rm M_\odot}$. 
Hence, our halo-to-disk mass fraction fits within this range.

Conroy et al. (2007) quotes a higher halo-to-stellar mass
ratio for the total stellar masses in excess of 
$10^{11}h^{-2}~{\rm M_\odot}$, using different definition for the halo
mass. We find that, after accounting for these differences, a factor
of 8 remains with their mean value, and a factor of 4 within their 
$1\sigma$. Interestingly, the quoted above results for the MW, will
differ by the same factor with Conroy et al. We do not find that
this difference will affect the processes discussed here, because
wew compare identical halos with and without baryons.

To summarize, in a number of associated publications, we have compared
the evolution of DM halos with and without baryons, from identical
initial conditions. Here we focus on two issues: assembly of the prime
halos and their relaxation processes. We find that baryons contribute
decisively to the evolution of the cusp region, with the baryon model
leading to the formation of an isothermal DM cusp, due to an adiabatic
contraction.  This cusp was ultimately dissolved by interactions with
the DM$+$baryon substructure and formed a flat density core (see also
Romano-Diaz et al.  2008b). The DM halo in this model nevertheless
remained more centrally concentrated compared to the pure DM
model. Futhermore, we find that the epoch of minor mergers is actually
dominated by interactions with the subhalos (with and without baryons).
These are responsible for ablating the cold, $< 3\times 10^4$~K,
gas component from the embedded disk, as well as heating up the
innermost halo gas and stars, causing their expansion out of the
region.  Heating the spheroidal gas, stellar and DM components in the
inner halo is the result of dynamical friction by the subhalos. The
disk, which started as gas dominated, becomes an intermediate Hubble
type by $z\sim 2$, and resembles that of the lenticular galaxies
after $z\sim 0.5$. The spheroidal-to-disk stellar mass ratio is
$\sim 0.7$ at this $z$ (within the disk radius), and the disk spiral 
activity has ceased accordingly, except during interactions with the 
subhalos.

Analyzing the halo assembly history we find that only a small, $\sim
11-15\%$, fraction of DM particles within \rts{} and $\sim 17-23\%$
within \rvmax, in PDM and BDM models respectively, are bound to within
these radii --- most of the DM particles perform much larger radial
excursions, thus mixing the smooth fraction of the inner halo
particles.  We compare this behavior with that of a non-singular
isothermal sphere in equilibrium. While the bound DM particles
assemble early within these radii, the fraction of bound particles
is small at all times. In other words, the mass distribution within
\rvmax\ is defined by DM particles which freely stream across this
region.
 
Lastly, the halos are only partially relaxed beyond
their virialization. Being an open system and accreting a
substantially inhomogeneous material (in all combinations of DM and
baryons), the degree of mixing in the halo is limited --- although
bound subhalos are tidally disrupted, their debris preserve
correlations in the phase-space over a few Gyrs time.  The mixing
process becomes even less efficient with time and phase-space
correlations (streamers) formed after $z\sim 1$ largely survive intact
to the present. Hence the halos are virialized but not `thermalized.'
This has interesting implications for the disk evolution and we
explore them in subsequent publications (Romano-Diaz et
al. 2008b,c, 2009a).  The persistence of substructure well past the
virialization assures that the trend toward a more complete violent
relaxation does not cease, but no attempt to quantify this process
here.

\acknowledgments 
Discussions with John Dubinski, Shardha Jogee and Simon White are 
gratefully acknowledged. We thank Ingo Berentzen for help with 
visualizations, and Jorge Villa-Vargas and Miguel A. Aragon-Calvo 
for helpful discussions on numerical issues.
This research has been partially supported by NASA/LTSA/ATP, NSF 
and STScI to I.S. STScI is operated by the AURA, Inc., under NASA 
contract NAS 5-26555. I.S. is grateful to the JILA Fellows for support.
Y.H. has been partially supported by the ISF (13/08).

%----------------------------------------------------------------------

%----------------------------------------------------------------------


\begin{thebibliography}{}
%\expandafter\ifx\csname natexlab\endcsname\relax\def\natexlab#1{#1}\fi
 
\bibitem[]{}Allgood, B., Flores, R.A., Primack, J.R., Kravtsov, A.V., Wechsler, R.H.,
      Faltenbacher, A., Bullock, J.S. 2006, MNRAS, 367, 1781

\bibitem[]{}Athanassoula, E., Misiriotis, A. 2002, \mnras, 330, 35

\bibitem[]{}Athanassoula, E. 2002, \apj, 569, L83 
 
\bibitem[]{}Bailin, J., Steinmetz, M. 2004, \apj, 616, 27 

\bibitem[]{}Barnes, E.I., Williams, L.L.R., Babul, A., Dalcanton, J.J.
     2007, \apj, 654, 814	
	 
\bibitem[]{}Barnes, J., Hut, P. 1986, Nature, 324, 446

\bibitem[]{}Berentzen, I., Shlosman, I. 2006, ApJ, 648, 807

\bibitem[]{}Bertschinger, E.\ 1987, \apjl, 323, L103 

\bibitem[]{}Bryan, G.L., Norman, M.L. 1998, \apj, 495, 80

\bibitem[]{}Bullock, J.S., Dekel, A., Kolatt, T.S., Kravtsov, A.V., Klypin,
   A.A., Porciani, C. \& Primack, J.R. 2001, \apj, 555, 240

\bibitem[]{}Bureau, M., Freeman, K.C., Pfitzner, D.W., Meurer, G.R. 1999,
      \aj, 118, 2158

\bibitem[]{}Christodoulou, D.M., Kazanas, D., Shlosman, I., Tohline, J.E.
      1995, \apj, 446, 472

\bibitem[]{}Conroy, C. et al. 2007, \apj, 654, 153    

\bibitem[]{}Debattista, V.P., Sellwood, J.A. 1998, \apj, 493, L5

\bibitem[]{}Dehnen, W. 2001, \mnras, 324, 273

\bibitem[]{}Dehnen, W. 2002, J. Comput. Phys., 179, 27 
      
\bibitem[]{}Dekel, A. et al. 2009, Nature, 457, 451

\bibitem[]{}Diemand, J., Kuhlen, M., Madau, P. 2007, \apj, 667, 859

\bibitem[]{}Diemand, J., Kuhlen, M., Madau, P., Zemp, M., Moore, B., Potter, D.,  
     Stadel, J., 2008, \nat, 454, 735

\bibitem[]{}Dubinski, J. 1992, \apj, 401, 441

\bibitem[]{}Dubinski, J. 1994, ApJ, 431, 617 

\bibitem[]{}Eisenstein, D.J. \& Hut, P. 1998, \apj, 498, 137 

\bibitem[]{}El-Zant, A. 2008, \apj, 681, 1058 

\bibitem[]{}Gao, L., De Lucia, G., White, S.D.M., Jenkins, A. 2004a, \mnras, 
     352, L1

\bibitem[]{}Gao, L., White, S.D.M., Jenkins, A., Stoehr, F., Springel, V. 
     2004b, \mnras, 355, 819

\bibitem[]{}Ghigna, S., Moore, B., Governato, F., Lake, G., Quinn, T., Stadel, J. 
    1998, \mnras, 300, 146

\bibitem[]{}Gnedin, O.Y., Kravtsov, A.V., Klypin, A.A., Nagai, D. 2004, \apj, 616, 16

\bibitem[]{}Hansen, S.H., Moore, B. 2006, New Astron., 11, 333

\bibitem[]{}Heller, C.H., Shlosman, I. 1994, \apj, 424, 84
 
\bibitem[]{}Heller, C.H., Shlosman, I., Athanassoula, E. 2007, ApJ, 671, 226  

\bibitem[]{}Hoffman, Y., Ribak, E. 1991, ApJ, 380, L5

\bibitem[]{}Hoffman, Y., Romano-Diaz, E., Shlosman, I., Heller, C.H. 2007, \apj,
     671, 1108

\bibitem[]{}Kazantzidis, S., Kravtsov, A.V., Zentner, A.R., Allgood, B., Nagai, D.,
    Moore, B. 2004, ApJ, 611, L73

\bibitem[]{} Kennicutt, R.C. 1998, \apj, 498, 541

\bibitem[]{}Keres, D., Katz, N., Weinberg, D.H., Dave, R. 2005, \mnras, 363, 2

\bibitem[]{}Klypin, A., Gottl\"ober, S., Kravtsov, A.V., Khokhlov, A.M.
     1999a, \apj, 516, 530

\bibitem[]{}Klypin, A., Kravtsov, A.V., Valenzuela, O., Prada, F. 1999b, \apj, 
    522, 82

\bibitem[]{}Li, Y., Mo, H.J., van den Bosch, F.C., Lin, W.P. 2007, \mnras,
     379, 689

\bibitem[]{}Lynden-Bell, D. 1967, \mnras, 136, 101

\bibitem[]{}Lynden-Bell, D., Kalnajs, A.J. 1972, \mnras, 157, L1 

\bibitem[]{}Martinez-Valpuesta, I., Shlosman, I., Heller, C.H. 2006, ApJ, 637, 
     214    

\bibitem[]{}McBride, J., Fakhougi, O., Ma, C.-P. 2009, \mnras, submitted; arXiv/0902.3659
    
\bibitem[]{}Moore, B., Ghigna, S., Governato, F., Lake, G., Quinn, T., Stadel, J., 
     Tozzi, P. 1999, \apj, 524, L19

\bibitem[]{}Naab, T., Johansson, P.H., Ostriker, J.P., Efstathiou, G. 2007, 
     \apj, 658, 710 

\bibitem[]{}Navarro, J.F., Frenk, C.S., White, S.D.M. 1997, ApJ, 490, 493 (NFW)   
    
\bibitem[]{}Ocvirk, P., Pichon, C., Teyssier, R. 2008, \mnras, 390, 1326

\bibitem[]{}Porciani, C., Dekel, A. \& Hoffman, Y. 2002, \mnras, 332, 325

\bibitem[]{}Reed, D., Governato, F., Quinn, T., Gardner, J., Stadel, J., Lake, G. 
      2005, \mnras, 359, 1537

\bibitem[]{}Romano-Diaz, E., Hoffman, Y., Faltenbacher, A., Heller, C.H., Jones, D.,
    Shlosman, I. 2006, ApJ, 637, L93  

\bibitem[]{}Romano-Diaz, E., Hoffman, Y., Heller, C.H., Faltenbacher, A., Jones, D.,
     Shlosman, I. 2007, \apj, 657, 56

\bibitem[]{}Romano-Diaz, E., Shlosman, I., Heller, C.H., Hoffman, Y. 2009a,
     in preparation (Paper~II)  

\bibitem[]{}Romano-Diaz, E., Shlosman, I., Hoffman, Y., Heller, C.H. 2008b, \apj,
      685, L105

\bibitem[]{}Romano-Diaz, E., Shlosman, I., Heller, C.H., Hoffman, Y. 2008c, \apj,
      687, L13 

\bibitem[]{}Shlosman, I. 2007, Pathways through an Eclectic Universe, J.H. Knapen, T.J. 
    Mahoney \& A.Vazdekis (eds.), ASP Conf. Ser. 390, p.~440, arXiv:0710.0630 

\bibitem[]{}Sparke, L.S. 2002, Disks of Galaxies: Kinematics, Dynamics \& Perturbations,
     ASP Conf. Proc., Vol. 275. (eds.) E. Athanassoula, A. Bosma \& R. 
    Mujica, 367 

\bibitem[]{}Spergel, D.N., et al. 2003, ApJS, 148, 175

\bibitem[]{}Spergel, D.N., et al. 2007, ApJS, 170, 377   

\bibitem[]{}Springel, V.,  Wang, J.,  Vogelsberger, M.,  Ludlow, A.,  Jenkins, A.,  
    Helmi, A.,  Navarro, J.F.,  Frenk, C.S., White, S.D.M. 2008, \mnras, 391, 1685

\bibitem[]{}Taylor, J.E., Navarro, J.F. 2001, \apj, 563, 483

\bibitem[]{}Tormen, G., Diaferio, A., Syer, D. 1998, \mnras, 299, 728
 
\bibitem[]{}Tremaine, S. \& Weinberg, M.D. 1984, \mnras, 209, 729 

\bibitem[]{}Wechsler, R.H., Bullock, J.S., Primack, J.R., Kravtsov, A.V., Dekel, A. 
    2002, \apj, 568, 52 

\bibitem[]{}Weinberg, D.H., Colombi, S., Davé, R., Katz, N. 2008, \apj, 678, 6

\bibitem[]{} van de Weygaert, R., \& Bertschinger, E.\ 1996, \mnras, 281, 84 

\bibitem[]{}White, S.D.M., Rees, M.J. 1978, \mnras, 183, 341 

\bibitem[]{}Xue, X.X. et al. 2008, \apj, 684, 1143

\bibitem[]{}Zait, A., Hoffman, Y., Shlosman, I. 2008, \apj, 682, 835 

\end{thebibliography}
\end{document}